%% file: main.tex
\documentclass[%
reprint,
superscriptaddress,
prd,
preprintnumbers,
nofootinbib,
 amsmath,amssymb,
 aps,floatfix
]{revtex4-1}

\renewcommand{\arraystretch}{1.2} 

\include{jpac}
\allowdisplaybreaks

\begin{document}

\title{Global analysis of charge exchange meson production at high energies} 

\author{J.~Nys}
\email{jannes.nys@ugent.be}
\affiliation{Department of Physics and Astronomy, Ghent University, Belgium}
\affiliation{Center for Exploration of Energy and Matter, Indiana University, Bloomington, IN 47403, USA}
\affiliation{Physics Department, Indiana University, Bloomington, IN 47405, USA}

\author{A.~N.~Hiller Blin}
\email{hillerbl@uni-mainz.de}
\affiliation{Institut f\"ur Kernphysik \& PRISMA Cluster of Excellence, Johannes Gutenberg Universit\"at, D-55099 Mainz, Germany}

\author{V.~Mathieu}
\affiliation{Theory Center, Thomas Jefferson National Accelerator Facility,
12000 Jefferson Avenue,  Newport News, VA 23606, USA}

\author{C.~Fern\'andez-Ram\'irez}
\affiliation{Instituto de Ciencias Nucleares, Universidad Nacional Aut\'onoma de M\'exico, Ciudad de M\'exico 04510, Mexico}

\author{A.~Jackura}
\affiliation{Center for Exploration of Energy and Matter, Indiana University, Bloomington, IN 47403, USA}
\affiliation{Physics Department, Indiana University, Bloomington, IN 47405, USA}

\author{A.~Pilloni}
\affiliation{Theory Center, Thomas Jefferson National Accelerator Facility,
12000 Jefferson Avenue,  Newport News, VA 23606, USA}

\author{J.~Ryckebusch}
\affiliation{Department of Physics and Astronomy, Ghent University, Belgium}

\author{A.~P.~Szczepaniak}
\affiliation{Center for Exploration of Energy and Matter, Indiana University, Bloomington, IN 47403, USA}
\affiliation{Physics Department, Indiana University, Bloomington, IN 47405, USA}
\affiliation{Theory Center, Thomas Jefferson National Accelerator Facility,
12000 Jefferson Avenue,  Newport News, VA 23606, USA}

\author{G. Fox}
\affiliation{School of Informatics and Computing, Indiana University, Bloomington, IN 47405, USA}

\collaboration{Joint Physics Analysis Center}

\preprint{JLAB-THY-18-2736}

 \begin{abstract}
Many experiments that are conducted to study the hadron spectrum rely on peripheral resonance production. Hereby, the rapidity gap allows the process to be viewed as an independent fragmentation of the beam and the target, with the beam fragmentation dominated by production  and decays of meson resonances. We test this separation by determining the kinematic regimes that are dominated by factorizable contributions, indicating the most favorable regions to perform this kind of experiments. In doing so, we use a Regge model to analyze the available world data of charge exchange meson production with beam momentum above $5\gev$ in the laboratory frame, that are not dominated by either pion or Pomeron exchanges. We determine the Regge residues and point out the kinematic regimes which are dominated by factorizable contributions.
 \end{abstract}

\maketitle

\section{\label{sec:introduction}Introduction}
\begin{figure}[htb]
\includegraphics[width=0.5\textwidth]{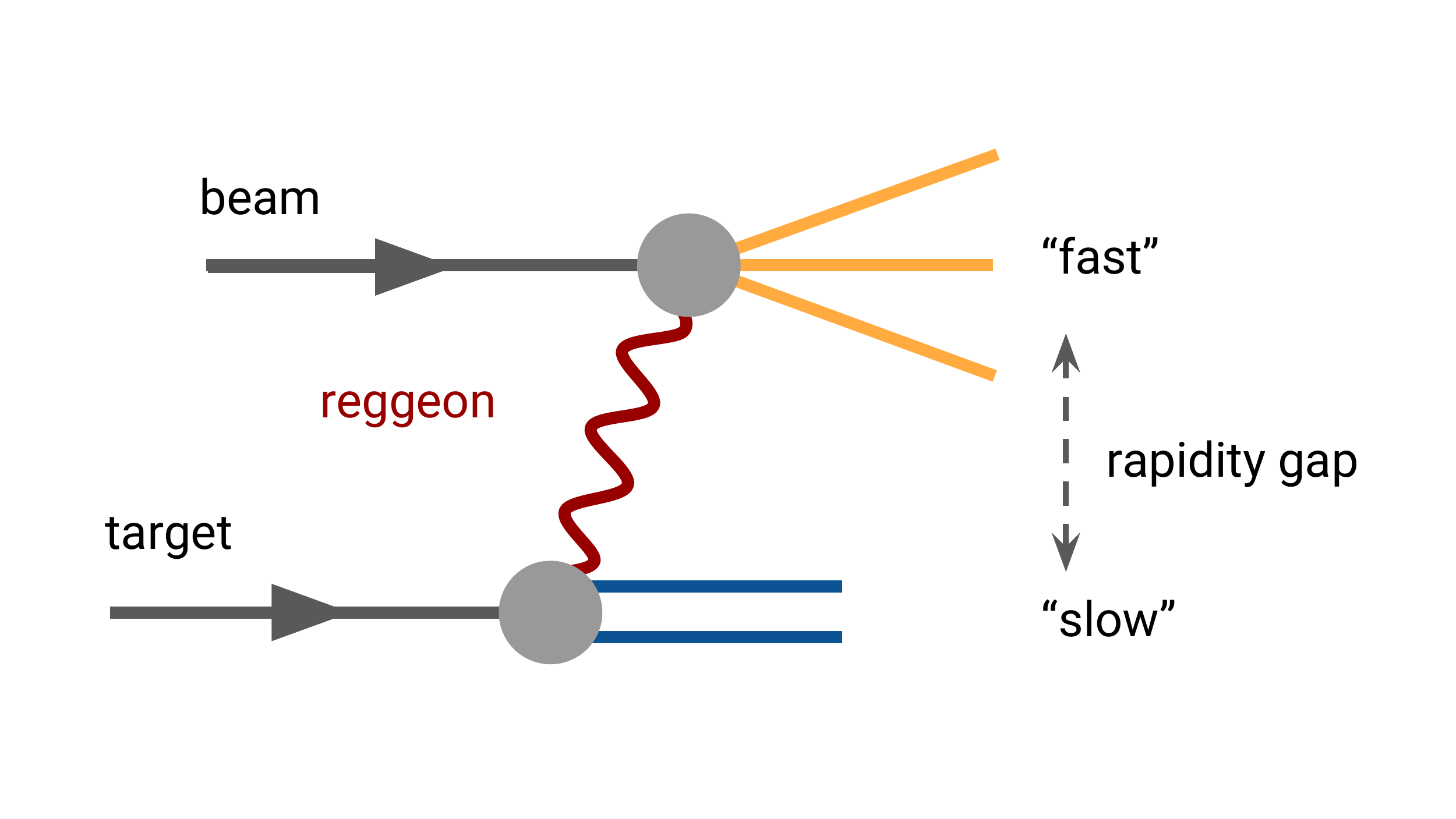}
\caption{
Illustration of factorization of peripheral meson production. Diffractive events are typically characterized by a gap in the rapidity distribution of the produced particles.\label{fig:diffraction}}
\end{figure}

The new generation of high statistics experiments 
{\it e.g.} Belle~II, BESIII, CLAS12, CMS, COMPASS, GlueX,  J-PARC,
LHCb, and \=PANDA,
have dedicated programs to study the hadron spectrum,
whose quantitative description is pivotal for 
a complete understanding of Quantum Chromodynamics (QCD).
These experiments demand a high level of precision in the 
amplitude analysis~\cite{Battaglieri:2014gca}
necessary to obtain reliable extractions of hadron properties from the data.
In particular, 
the diffraction of photons or mesons on the nucleon target
at high energies, as studied at GlueX, CLAS12 and COMPASS, 
is expected to provide
information on hybrids, 
exotics and the gluonic degrees of freedom,
via independent fragmentation of the beam and of the target (see Fig.~\ref{fig:diffraction}), 
with the beam fragmentation dominated by production and decays of mesons.

Regge phenomenology underlies such processes and provides the theoretical framework for studying high energy scattering. In Regge theory, resonances in the exchanged channel are related to each other and are described by the Regge trajectories, also referred to as reggeons. Specifically, the pion diffractive dissociation at COMPASS is dominated by exchanges of reggeons with vacuum quantum numbers, including the Pomeron ($\pomeron$). Photon induced reactions at the Jefferson Lab (JLab) may also proceed by exchange of reggeons with non-vacuum quantum numbers. Regge trajectories provide specific information about the dynamics responsible for the formation of resonances~\cite{Londergan:2013dza,Fernandez-Ramirez:2015fbq,Pelaez:2017sit} that can be used to constrain amplitudes of other reactions, \eg production of light hadrons in heavy flavor hadron decays. In order to confidently separate the beam and target fragmentation in peripheral scattering, it is necessary to establish the validity of Regge pole factorization. 
  Establishing the production mechanism in production of  resonances is also the necessary first step in determination of their quantum numbers and other characteristics, as their quark model nature~\cite{Mandula:1970wz}. This is particularly relevant when searching for new states, \eg hybrid mesons, which is one of the main goals of the spectroscopy program at JLab~\cite{AlGhoul:2017nbp,Glazier:2015cpa}.
  
 In the near future new data on peripheral resonance production will be coming primarily from JLab experiments, and therefore it is important to validate Regge mechanisms for beam energies of $E_\gamma \sim O(10\gev)$.
 Even though high energy peripheral processes are expected to be dominated  by exchanges of leading Regge poles, there are sub-leading singularities, \eg Regge cuts and/or poles in daughter trajectories, which have to be assessed~\cite{Collins:1977jy}. 
With this goal in mind, we have recently studied $\pi N$ scattering, and $\pi^0 N$, $\eta N$, $\pi \Delta$ and neutral vector meson  photoproduction~\cite{Mathieu:2015gxa,Mathieu:2015eia,Nys:2016vjz,Nys:2017xko,Mathieu:2017but,Mathieu:2018xyc,Mathieu:2017jjs}, obtaining several results that we briefly  summarize below. 
  In $\pi^0$ photoproduction, we used finite energy sum rules (FESR's)~\cite{Mathieu:2017but} to demonstrate that there is a good agreement between the partial wave models (PWA) for low energy amplitudes~\cite{Kamano:2013iva,Anisovich:2011fc,Anisovich:2012ct,Ronchen:2014cna,Ronchen:2015vfa,Chiang:2001as,Workman:2012jf,Briscoe:2012ni}, and the Regge parametrization of the 
  high-energy data. In $\eta$ photoproduction, the PWA models are less constrained by the available low-energy data and we have shown how the high-energy data can help reducing uncertainties, specifically those related to unnatural exchanges~\cite{Mathieu:2017but,Nys:2016vjz}.
   Our prediction for the $\pi^0$ photoproduction cross section~\cite{Mathieu:2015eia} compares favorably with the CLAS data~\cite{Kunkel:2017src}, and our  results on $\pi^0$, $\eta$ and $\eta'$ photoproduction beam asymmetries~\cite{Mathieu:2018xyc,Mathieu:2017jjs} are in agreement with the  GlueX results~\cite{AlGhoul:2017nbp,DNP2017}. The main conclusion one can draw from these comparisons is that, at forward scattering angles, the natural Regge poles dominate over the unnatural ones and over the non-pole contributions. 
 In general, for natural exchanges we have found a good agreement with factorization. Specifically, 
a zero in the residue of the $\omega$ exchange in $\pi N$~\cite{Mathieu:2015gxa}  implies a similar behavior for the photoproduction reactions with $\omega$ exchange. Indeed, 
 a zero is found in the $\eta$ photoproduction amplitude and a strong dip is present in the cross section for $\pi^0$ photoproduction~\cite{Nys:2016vjz}. 
Complementary to these analyses, we now study reactions with meson beams sharing the same nucleon residues. For these reactions, the amount of high-energy data is abundant, which allows for a detailed study of both the residue factorizability, and the energy dependence of the observables. Additionally, mesonic beams allow for less exchanges compared to photon beams, therefore allowing us to study the dominant natural exchanges in isolations. For those kinematics where Regge factorization holds, information about the residues can be applied to photoproduction reactions.

Since the Regge picture has been well established for
$\pi N$ scattering and $\pi$ and $\eta^{(\prime)}$ photoproduction off the proton, in this work, we proceed to examine the Regge pole model in a global analysis of several quasi-two-body reactions of interest to peripheral resonance production.   When sub-leading contributions, such as Regge cuts or daughters, are accounted for in the amplitudes, the factorization approximation is violated. We aim to identify the kinematics for which such violations can be expected, while for processes dominated by factorizable exchanges we provide amplitudes and residues that are compatible with the world data at high energies. These can be used to model the production mechanism in fragmentation experiments, allowing the isolation of the resonant part intended to search for hybrids.

In the Regge pole approximation, the amplitudes are well constrained by unitarity and analyticity and are specified by a small number of parameters. Thus, in principle, a large enough data set in principle makes it possible to test the Regge pole dominance hypothesis~\cite{Fox:1967zza}. In this work we perform a global analysis of all available data on charge exchange (CEX) quasi-two-body reactions with meson beams, that are dominated by vector and tensor Regge trajectories. Except for the Pomeron exchange, which does not contribute to CEX reactions, vector and tensor exchanges are expected to dominate in the energy range of interest. Furthermore, we exclude processes in which pion exchange is possible. At high energies, the pion pole is close to the physical region and becomes more sensitive to the sub-leading Regge contributions. This, in general, requires a special treatment~\cite{piexchangeFox,vincent_fesr_pion,Nys:2017xko}. The data set considered in this paper includes $23$ reactions and $1271$ differential cross section data points, as described in Section~\ref{sec:results}, and summarized in Table~\ref{tab:data_refs}.

The paper is organized as follows. We discuss the main features of the formalism in Sections~\ref{sec:formalism} and~\ref{sec:constraints}, leaving the technicalities to the Appendices. The results of the fits are discussed in Section~\ref{sec:results} and in Section~\ref{sec:conclusions} we summarize the main conclusions.
The kinematics and our conventions are discussed in detail in Appendix~\ref{sec:conventions}. Appendix~\ref{sec:factorization} contains a summary of the effect of factorization of the Regge residues on the forward behavior of the helicity amplitudes. The interaction Lagrangians used in the fits are contained in Appendix~\ref{sec:lagrangians}, with estimates of the corresponding coupling constants derived in Appendix~\ref{sec:couplings_estimations}. Our method for building Regge amplitudes from single-particle exchange amplitudes is discussed with an example in Appendix~\ref{sec:amplitudes}. Finally, Appendix~\ref{sec:rotation} provides the expressions that allow one to determine the $t$-channel helicity residues directly from the $s$-channel residues and {\it vice versa}, without the need for introducing Lagrangians.

\section{\label{sec:formalism}Formalism}
\begin{figure}[tbh]
\centering
\includegraphics[width=0.2\textwidth]{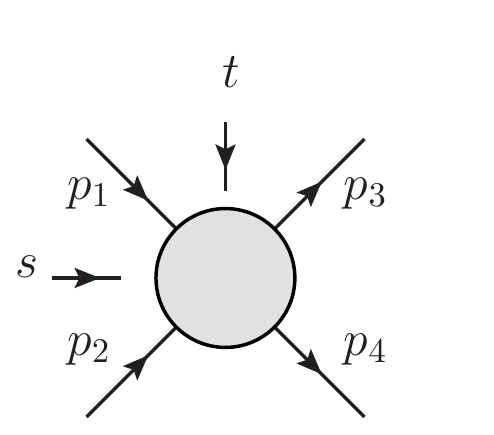}%
\includegraphics[width=0.3\textwidth]{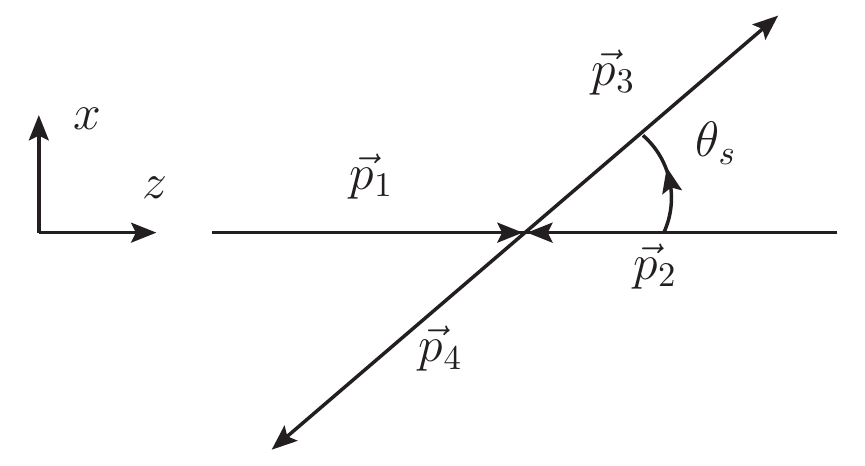}
\caption{Kinematics for the reaction $1 + 2 \to 3 + 4$ in the $s$-channel center of mass frame.\label{fig:kinematics_schematic}}
\end{figure}

We consider reactions of the type (see Fig.~\ref{fig:kinematics_schematic})
\begin{align}
1(p_1,\mu_1) + 2(p_2,\mu_2) \to 3(p_3,\mu_3) + 4(p_4,\mu_4)\,,
\end{align}
where the $p_i$'s are the 4-momenta and the $\mu_i$'s are the helicities in the center of mass frame, referred to as the $s$-channel frame. The standard Mandelstam variables are $s = (p_1+p_2)^2$, $t = (p_1-p_3)^2$ and $u=(p_1-p_4)^2$.
We parametrize the high-energy $s$-channel helicity amplitudes following the analysis of Cohen-Tannoudji~\etal\cite{Cohen-Tannoudji:1968eoa} which leads to a  factorized form as discussed in~\cite{Fox:1967zza}. Specifically, in the large $s$ limit the amplitude of a Regge pole ($e$) described by a trajectory $\alpha_e(t)$, which in the $s$-channel physical region is approximated by a linear function, $\alpha_e(t) = \alpha_e^0 + \alpha_e^1 t$, is given by  
\begin{align}\label{eq:most_general_regge_form}
A_{\mu_4 \mu_3 \mu_2 \mu_1}(s,t) =& - \frac{\pi \alpha_e^{1}}{2} \beta^{e}_{\mu_4 \mu_3 \mu_2 \mu_1}(t) \nonumber \\
& \times \frac{\zeta_e+ e^{-i \pi \alpha_e(t)}}{\sin \pi \alpha_e(t)} \left(\frac{s}{s_0}\right)^{\alpha_e(t)} \,,
\end{align}
were $\zeta_e = \pm 1$ is the reggeon signature and $s_0$ is a  scale above which the pole approximation is expected to dominate, $s \gtrsim s_0$. The latter is related to the range of the strong interaction. In the following we use $s_0 = 1\gev^2$. We refer to right (wrong) signature points as those kinematics for which the signature factor $\zeta_e+ e^{-i \pi \alpha_e(t)}$ in Eq.~\eqref{eq:most_general_regge_form} is finite (vanishes).
Except for kinematic factors, discussed below, Regge theory does not fix the $t$-dependence of the residues $\beta^{e}_{\mu_4 \mu_3 \mu_2 \mu_1}(t)$. Unitarity in the $t$-channel requires that the trajectory and residues are real functions in the kinematic region of interest, and that the $t$-channel helicity residues are factorizable in a separate contribution coming from the meson vertex and a contribution from the baryon vertex. It was shown in~\cite{Fox:1967zza} that in the high-energy limit, due to  properties of the crossing matrix, the $s$-channel helicity residues must also be factorizable. 
Hence, the residues $\beta^{e}_{\mu_4 \mu_3 \mu_2 \mu_1}(t)$ in Eq.~\eqref{eq:most_general_regge_form} can be written in a product form~\cite{Gribov:2009zz}
\begin{align}\label{eq:factorization}
\beta_{\mu_4 \mu_3 \mu_2 \mu_1}^e(t) = \beta^{e 2 4}_{\mu_2 \mu_4}(t)\,  \beta_{\mu_1 \mu_3}^{e 1 3}(t) \,.
\end{align}
In the following, the $\beta^{e 2 4}_{\mu_2  \mu_4}(t)$  and $\beta^{e 1 3}_{\mu_1 \mu_3}(t)$ will be referred to as the bottom and top or meson and nucleon residues respectively. The superscripts indicate 
which external particles the residue depends upon. This factorized form illustrates the role played by coupled channels, since the residues $\beta^{e i j}_{\mu_i  \mu_j}(t)$ in Eq.~\eqref{eq:factorization} (where $(i,j) = (1,3)$ or $(i,j) = (2,4)$) are shared among various reactions which involve the same particles.

The function $\sin\pi\alpha_e(t)$ in the denominator of   Eq.~\eqref{eq:most_general_regge_form} reflects the existence of particle poles in the $t$-channel with integer spins $J$ given by the value of the trajectory at a pole, $J=\alpha_e(t=m_e^2)$. Regge amplitudes involve particles of definite parity and  naturality $\eta = P(-1)^J = \pm 1$, which is accounted for by the signature factor $(\zeta_e + e^{-i\pi\alpha_e(t)})/2$. For physical spins, it reduces to $(\zeta_e  + (-1)^J)/2$ and retains  only the poles which satisfy $\eta P = \zeta_e$. Since poles with negative spins are unphysical, the residues must contain additional zeros. For example, zeros in even (odd) signature  residues must be present at values of $t$ for which $\alpha_e(t)$ is equal to an  even (odd) negative integer. In addition $\alpha_e=0$ in the even signature trajectory may occur at a value of $t$ in the $s$-channel physical region. For some trajectories, this would correspond to an unphysical exchange of a ghost spin-0 particle. This happens in the $a_2$ trajectory\footnote{We denote reggeons by their lowest spin particles. Also, $K^* \equiv K^*(892)$ and $K^*_2 \equiv K^*_2(1430)$.}, where $\alpha_{a_2}(t) = 0$ corresponds to negative values of $t$, and therefore an additional zero must be included at $\alpha_{a_2}(t) = 0$ in the residue of the $a_2$ exchange. Hadron Regge trajectories are observed to satisfy the so-called  exchange degeneracy (EXD), which can be understood in terms of local Regge-resonance duality and large-$N_c$ limit~\cite{Mandula:1970wz,Harari:1981nn}. As a consequence, residues of the $a_2$ and $\rho$ exchanges should have the same $t$-dependence\footnote{Two types of EXD can be distinguished: \textit{weak} EXD, which requires the trajectories of opposite signature exchanges to be equal, and \textit{strong} EXD, which additionally requires the residues to be equal. We will use the term EXD interchangeably for both cases, since the type of EXD will be clear from the context.}. 
 Together with the requirement for zeros at the negative values of $\alpha_e$ with the right signature, the residues of $a_2$ and $\rho$ must contain zeros at all non-positive spins. While the above reasoning invokes EXD, which holds only approximately, the zeros at $\alpha_e(t) = 0$ are more general, as will be discussed at the end of this Section. Zeros at non-negative opposite signature spins, such as the zero at $\alpha_\rho(t) = 0$, are referred to as wrong signature zeros (WSZ). To include all the expected zeros, as discussed above we require~\cite{Collins:1977jy,Fox:1973by,Irving:1977ea}
\begin{align}\label{eq:NWSZ_beta}
\beta_{\mu_4 \mu_3 \mu_2 \mu_1}^e(t) \propto \frac{1}{\Gamma(\alpha_e(t) - l_e+1)} \,,
\end{align}
where $l_e$ is the lowest physical spin on the trajectory $\alpha_e(t)$, or on the trajectory of its exchange-degenerate partner, \ie $l_\rho = l_{a_2} = 1$.
Since it is not known {\it a priori} which one of the two factorizable residues should the zeros be attributed to, we pull this factor out of the product of residues and absorb it in the definition of the remaining terms in Eq.~\eqref{eq:most_general_regge_form}. 
The trajectories are assumed to be linear functions and are given in Table~\ref{tab:trajectories}. Notice that we assume weak degeneracy for all exchanges: $\alpha_\rho(t) = \alpha_{a_2}(t)$ and $\alpha_{K^*}(t) = \alpha_{K^*_2}(t)$.

As discussed in Appendix~\ref{sec:factorization}, for $t \to 0$ the most singular behavior of residues corresponding to $t$-channel exchanges of definite parity is given by $\beta^{e i j}_{\mu_i \mu_j} (t)\sim \sqrt{-t}^{\abs{\mu_i - \mu_j}}$. In order to make this kinematic $t$-dependence explicit, we define the reduced residues $\betahat_{\mu_i \mu_j}(t)$, 
\begin{align}\label{eq:sqrtt_factors}
\beta_{\mu_i \mu_j}^{eij}(t) = \sqrt{-t}^{\abs{\mu_i - \mu_j}} \hat\beta^{eij}_{\mu_i \mu_j}(t) \,,
\end{align}
which are regular in $t$. 
The general form of the large $s$ amplitude implied by Regge theory is therefore given by
\begin{align}
\label{eq:asymptotic_regge_form} 
A_{\mu_4 \mu_3 \mu_2 \mu_1}(s \gg s_0,t) =& \sqrt{-t}^{\abs{\mu_1 - \mu_3}} \sqrt{-t}^{\abs{\mu_2 - \mu_4}} \nonumber \\
& \times  \betahat^{e 1 3}_{\mu_1 \mu_3}
(t) \betahat^{e 2 4}_{\mu_2 \mu_4}(t) \mathcal{F}_e(s,t) \,,
\end{align}
where $\mathcal{F}_e(s,t)$ is defined below.
In the phenomenological analysis of the data, we have found 
 that one obtains better fits if the $t$-dependence implied by the exact angular behavior of the amplitude is used instead of its limit corresponding the high energy scattering in the  forward direction, $z_s \to 1+2t'/s$ where $z_s$ is the cosine of the scattering angle in the $s$-channel frame, and $t' = t-t_{min}$ with $t_{min} = t(z_s=+1) =O(1/s)$~\cite{Irving:1977ea}.
Specifically, from the overall angular-momentum conservation it follows that  the $s$-channel helicity amplitude is proportional to the half-angle factor~\cite{Mikhasenko:2017rkh,Pilloni:2018kwm}
\begin{align}\label{eq:half_angle_factor}
\xi_{\mu \mu'}(s,t) = \left(\frac{1-z_s}{2}\right)^{\abs{\mu - \mu'}/2}\left(\frac{1+z_s}{2}\right)^{\abs{\mu + \mu'}/2} \,,
\end{align}
where $\mu = \mu_1 - \mu_2$ and $\mu' = \mu_3- \mu_4$ are the net helicity in the initial and final state, respectively.  
The half-angle factor incorporates the kinematic singularities in $t$ of the $s$-channel helicity amplitude, and in the forward direction at high energies it reduces to 
\begin{align}\label{eq:half_angle_asy}
\xi_{\mu\mu'}(s,t)\xrightarrow[]{s\to\infty}\sqrt{\frac{-t}{s}}^{\abs{\mu-\mu'}} \,.
\end{align}
By comparing with the asymptotic limit of the Regge pole expression given by Eq.~\eqref{eq:asymptotic_regge_form}, we obtain 
\begin{eqnarray}
& & A_{\mu_4\mu_3\mu_2 \mu_1} = \frac{\xi_{\mu \mu'}(s,t)}{ \sqrt{\frac{-t}{s}}^{\abs{\mu - \mu'}}  }  \nonumber\\ && \times  \left[\sqrt{-t}^{\abs{\mu_1 - \mu_3}} \sqrt{-t}^{\abs{\mu_2 - \mu_4}} \betahat^{e 1 3}_{\mu_1 \mu_3}(t) \betahat^{e 2 4}_{\mu_2 \mu_4}
(t) \mathcal{F}_e(s,t)\right] \,, \nonumber \\
\end{eqnarray}
where the half-angle factor is fully taken into account. 
Alternatively we can restore the half angle factor 
 by multiplying the Regge formula (Eq.~\ref{eq:asymptotic_regge_form}) 
by a factor $\mathcal{R}$
\begin{align}
\mathcal{R}(s,t) \equiv \left( \frac{1 - z_s}{2} \frac{\nu}{-t} \right)^{\frac{1}{2} \abs{\mu - \mu'}} \left(\frac{1 + z_s}{2} \right)^{\frac{1}{2} \abs{\mu + \mu'}} \,,
\end{align}
where $\nu = (s-u)/2$. 
It is worth noting that the small $\abs{t}$ behavior imposed by angular-momentum conservation $\beta_{\mu_4 \mu_3 \mu_2 \mu_1}^e(t) \propto \sqrt{-t}^{\abs{(\mu_1 - \mu_3) - (\mu_2 - \mu_4)}}$ in Eq.~\eqref{eq:half_angle_asy} is weaker than the behavior $\beta_{\mu_4 \mu_3 \mu_2 \mu_1}^e(t) \propto \sqrt{-t}^{\abs{\mu_1 - \mu_3}+\abs{\mu_2 - \mu_4}}$ introduced by the additional requirement of factorization of the residue in Eqs.~\eqref{eq:sqrtt_factors} and \eqref{eq:asymptotic_regge_form}. 
 We have normalized $\mathcal{R}$ in such a way that $\mathcal{R} \to 1$ for $s\to \infty$, such that the $s$-dependence of the helicity amplitude remains $s^{\alpha_e(t)}$, as in Eq.~\eqref{eq:asymptotic_regge_form}. The final functional form for the amplitudes used for the fits reads
\begin{align}\label{eq:general_regge_form} 
A_{\mu_4 \mu_3 \mu_2 \mu_1}(s,t) =& \mathcal{R}(s,t)\sqrt{-t}^{\abs{\mu_1 - \mu_3}} \sqrt{-t}^{\abs{\mu_2 - \mu_4}}  \nonumber \\
&\times \betahat^{e 1 3}_{\mu_1 \mu_3}(t) \betahat^{e 2 4}_{\mu_2 \mu_4}(t) \mathcal{F}_e(s,t) \,.
\end{align}
Regge theory does not predict the full $t$-dependence and the single-particle model only approximates the $t$-dependence close to the single particle pole. An exponential factor is introduced with a slope parameter and in the following we will implicitly assume the presence of an exponential factor in the residues
\begin{align}\label{eq:general_damping}
\betahat^{eij}_{\mu_i \mu_j}(t) \propto e^{b^{eij}_{\mu_i \mu_j} t} \,.
\end{align}

The function $\mathcal{F}_e(s,t)$ is often referred to as the Regge propagator, 
and is given by
\begin{align}\label{eq:regge_propagator}
\reggepropagator{e}(s,t) &= -\frac{\zeta_e \pi \alpha_e^{1}}{\Gamma(\alpha_e(t) - l_e+1)} \frac{1+\zeta_e e^{-i \pi \alpha_e(t)}}{2 \sin \pi \alpha_e(t)} \left(\frac{s}{s_0}\right)^{\alpha_e(t)}\,.
\end{align}
In the $t \to m_e^2$ limit, where $m_e$ is the mass of the meson with lowest spin $J_e$ on the trajectory $\alpha_e(t)$, one recovers the known single-particle propagator\footnote{Note that our convention is different than the one in~\cite{Irving:1974ak} in the $\zeta_e$ prefactor. Our convention is correctly normalized to the single particle pole as in Eq.~\eqref{eq:prop_limit}, while the definition of~\cite{Irving:1974ak} flips sign according to the signature of the exchange. Naturally, this normalization affects the sign of the residues of the negative signature exchanges ($\zeta_e = -1$). We absorb the sign difference into the top vertices of those exchanges.} 
\begin{align}\label{eq:prop_limit}
\reggepropagator{e}(s,t) \underset{t\to m_e^2}{\longrightarrow} \frac{(s/s_0)^{J_e}}{m_e^2 - t}\,.
\end{align}
This property illustrates that the Regge amplitude in Eq.~\eqref{eq:general_regge_form} reduces to the single-particle exchange amplitude near the particle poles. This property constrains the residues near the pole to be numerically close to the phenomenological values from single-particle exchange models~\cite{Irving:1977ea}.

Note that in our approach the reggeons are used to  directly construct the $s$-channel helicity amplitudes. A more natural approach would be to start in the rest frame of the reggeon, i.e.\ in the $t$-channel center-of-mass (c.m.) frame and apply crossing relations.  
 In the $t$-channel frame the relation between the residue zeros and angular momentum conservation is more transparent. 
  Consider for example, the $t$-channel amplitude  $\pi \pi \to \overline{N} N$ with a $\rho$ exchange. A spin-$0$ exchange is unphysical in the $t$-channel helicity flip component. Such a contribution is referred to as `nonsense'~\cite{Collins:1977jy}. 
  Therefore, a nonsense wrong-signature zero (NWSZ) is often introduced at $\alpha_\rho=0$ \textit{only} in the $t$-channel helicity-flip component, in order to remove such a spurious contribution.
For the $t$-channel non-flip component, the Finite-Energy Sum Rule analysis indeed show a finite residue at the wrong-signature point ($\alpha_\rho = 0$)~\cite{Mathieu:2015gxa}.
However, starting from the $t$-channel, a global analysis is tedious and less straightforward, as discussed for example in~\cite{Mikhasenko:2017rkh}. Indeed, for each channel one must trace the kinematic singularities in $t$ and remove them in order to construct a set of helicity amplitudes which contains only dynamic singularities in $t$. In contrast, the $t$-singularities of the $s$-channel helicity amplitudes are easy to find  as they originate entirely from the half-angle factors defined in Eq.~\eqref{eq:half_angle_factor}, see Appendix~\ref{sec:conventions}. By merely dividing out the half-angle factors, one is able to write down $s$-channel helicity amplitudes that are free from kinematic singularities in $t$. For a global phenomenological analysis it is therefore preferential to deal directly with the $s$-channel amplitudes. 

\section{Constrained SU(3)-EXD fit}
The total number of parameters describing Regge residues is substantial. In order to obtain a robust estimate we first carry out a fit to the data in which we impose both SU(3) and exchange degeneracy (EXD). Furthermore, we fix the $t$-dependence of the reduced residues $\betahat^{eij}_{\mu_i \mu_j}$ (see Eq.~\ref{eq:general_regge_form}) using a single-particle exchange model obtained from the effective Lagrangians as discussed in  Appendices~\ref{sec:lagrangians} and~\ref{sec:amplitudes}. We refer to this analysis as the `global SU(3)-EXD fit', which is similar to the analyses of~\cite{Irving:1977ea,Fox:1973by}. The results are shown in Section~\ref{sec:constrained_fit}. We later relax all these assumptions for a second {\it unconstrained} fit, in Section~\ref{sec:unconstrained_fit}.
 
\subsection{Constraints\label{sec:constraints}}
To derive the SU(3) couplings, we consider SU(3) relations based on the Lagrangians in Appendix~\ref{sec:lagrangians} for the single-particle exchange model. Using the relation between the single-particle and the Regge residues one obtains the SU(3) constraints for the latter, using the residues of the $\rho$ and $K^*$ exchanges\footnote{We denote reggeons by their lowest spin particles. Also, $K^* \equiv K^*(892)$ and $K^*_2 \equiv K^*_2(1430)$.} as input.
Since we do not consider reactions dominated by $\pomeron$ or $\pi$ exchange, we do not fit $NN$ cross section data. If we did, it would allow us to constrain the overall normalization of the fit, through the determination of $\betahat^{eNN}$. Instead, we extract the top vertices from resonance decay widths. 
We fix $\betahat^{\rho \pi^- \pi^-} = 8.4$ using the Lagrangian coupling, $g_{VPP} = -4.2$, as discussed in Appendix~\ref{sec:couplings_estimations}. Note that the sign is chosen such that the contributions to the total cross sections match correctly.   
While SU(3) allows one to relate the various Regge residues of vector exchanges, it does not relate these to residues of tensor exchanges ($a_2$ and $K^*_2$). The $a_2$ and $K^*_2$ couplings for the global SU(3)-EXD fit are obtained by demanding EXD for the helicity residues. In summary, we use the following relations obtained from duality arguments~\cite{Mandula:1970wz}
\begin{subequations}
\begin{align}
\betahat^{\rho K^+ K^+} &= -\betahat^{a_2 K^+ K^+} \,, \label{eq:EXD_constr1} \\
\betahat^{\rho p p} &= \betahat^{a_2 p p} \,, \label{eq:EXD_constr2} \\ 
\betahat^{\rho p \Delta^+} &= \betahat^{a_2 p \Delta^+} \,,  \label{eq:EXD_constr3}
\end{align}
\end{subequations}
for any helicity combination. Since $K^+ p \to K^+ p$ and $K^+ n \to K^+ n$ are exotic in the $s$-channel, duality requires that $\betahat^{\rho K^+ K^+} \betahat^{\rho p p} = -\betahat^{a_2 K^+ K^+} \betahat^{a_2 p p}$, since the propagator is normalized according to Eq.~\eqref{eq:prop_limit} (and similarly for $f$ and $\omega$ exchanges). Note that exact EXD is not necessarily fulfilled within the single-particle model. In particular, EXD requires the residues of two exchanges to be equal for any $t$, which might not be possible, especially when only a subset of the interaction Lagrangians is considered, as is often the case in the literature. Therefore, an `EXD propagator' (which amounts to adding or subtracting the propagators of EXD contributions) is usually introduced to circumvent this issue, while strong EXD is in fact a property of the residues.
In the latter approach, one effectively assumes that both exchanges have the single-particle residue of the lowest spin exchange.

It is worth noting that EXD is sometimes used incorrectly.
It is interpreted as a property of a single reggeon, while it is in fact a relation  between different Regge pole contributions.  Therefore, the EXD propagator can only be used in reactions where both EXD partners are allowed. For instance, there is no strong EXD for the residues $\beta^{e \gamma \pi}_{\mu_\gamma \mu_\pi}(t)$ of $\pi$ photoproduction reactions, since there is no exotic $s$-channel reaction containing these residues.

In the global SU(3)-EXD fit, we opt to introduce a single exponential damping factor for each independent helicity configuration $b_{\textrm{nf}}$, $b_{\textrm{sf}}$, with nf and sf referring  to helicity non-flip and single flip amplitudes. These are fitted to the data. In this fit we set the double-flip amplitudes to zero. 

\subsection{Results\label{sec:results}}\label{sec:constrained_fit}
In the following, we present results of the global analysis of the high-momentum ($\plab \geq 5\gev$) differential cross section data for a large number of reactions, including reactions with $\pi$ and $K$ beams that are dominated by $\rho$, $a_2$, $K^*$ and $K^*_2$ exchanges\footnote{We do not report fits for tensor meson production, since the scarceness of the data strongly hinders a reliable fit.}. Hereby, we consider the channels with strangeness and charge exchange (CEX). The data used in this analysis are listed in Table~\ref{tab:data_refs}. 

\begin{figure*}[h]
\centering
\begin{subfigure}{.5\textwidth}
\centering
\includegraphics[width=\textwidth]{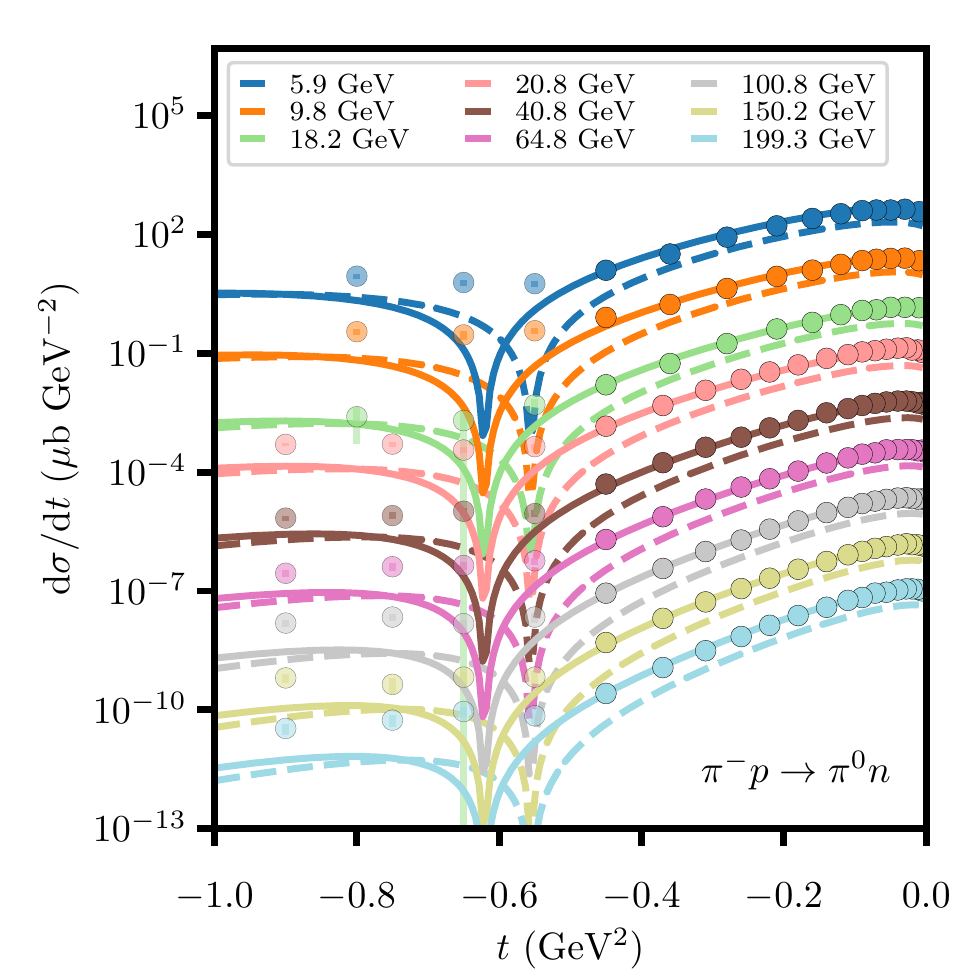}
  \caption{$\pi^- p \to \pi^0 n$}
  \label{fig:PI-_P_PI0_N}
\end{subfigure}%
\begin{subfigure}{.5\textwidth}
\centering
\includegraphics[width=\textwidth]{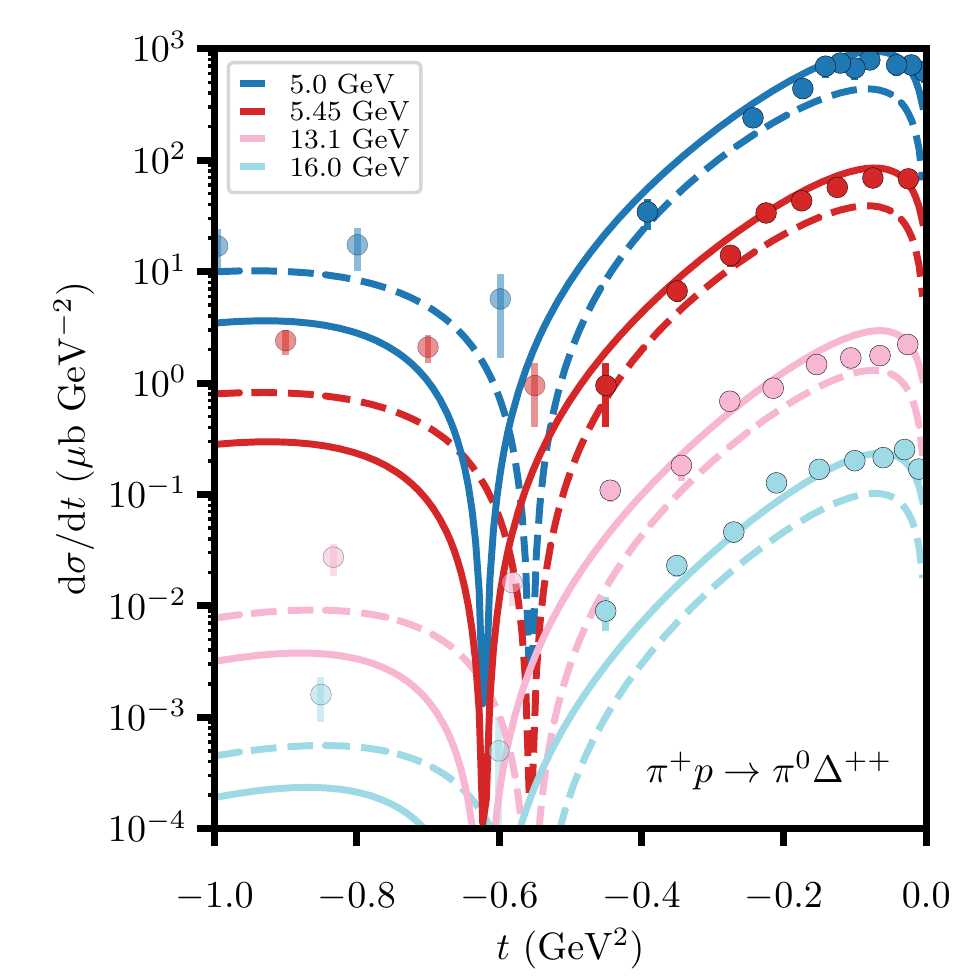}
  \caption{$\pi^+ p \to \pi^0 \Delta^{++}$ \label{fig:PI+_P_PI0_DELTA_1232P33_++}}
\end{subfigure}%
\caption{Differential cross section for the channels dominated by $\rho$ exchange. Dashed (solid) lines represent the global SU(3)-EXD (unconstrained) fit. The cross section for the  lowest $\plab$ has not been rescaled. For momentum value $i$ (in ascending order, with $i=0$ the lowest $\plab$), the cross section has been rescaled by a factor $(0.1)^i$. The legend shows the $\plab$ values of the measurements. The transparent data set has not been included in the unconstrained fit, but are shown for completeness.\label{fig:PI_PI_RHO}}
\end{figure*}

\begin{figure*}[h]
\begin{subfigure}{.5\textwidth}
\centering
\includegraphics[width=\textwidth]{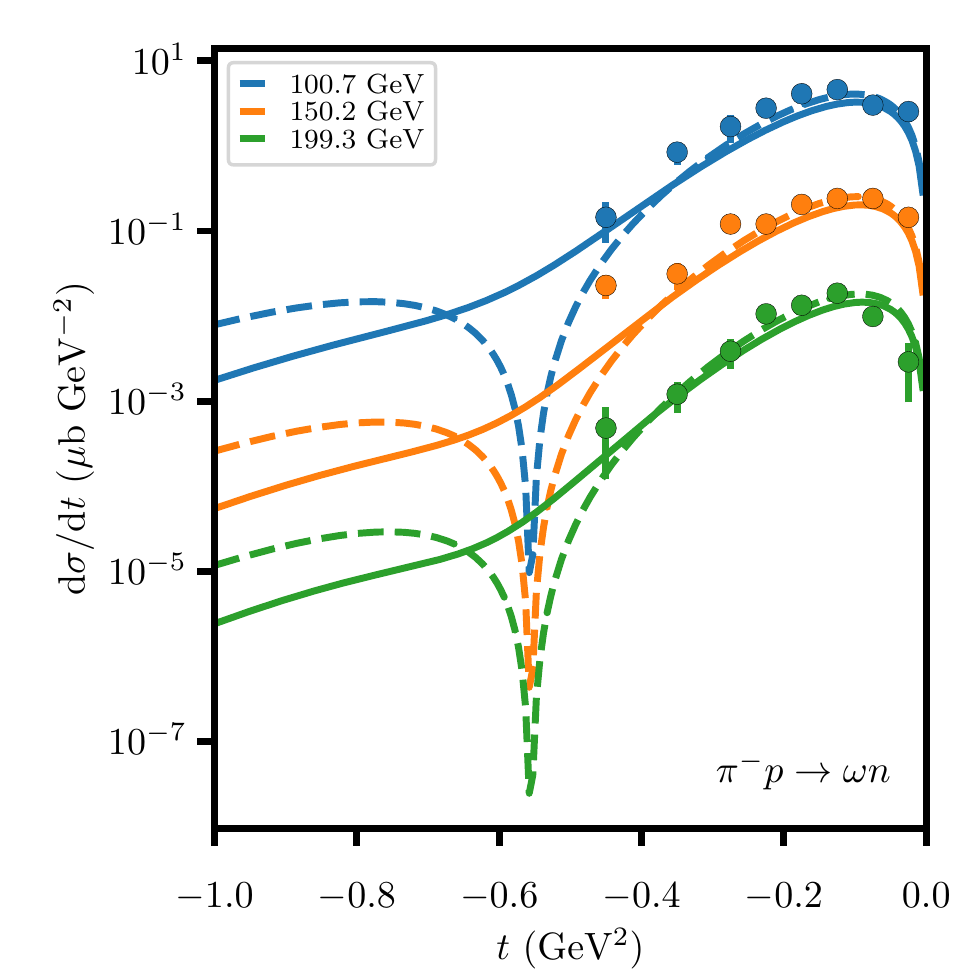}
  \caption{$\pi^- p \to \omega n$, $\plab > 100\gev$\label{fig:PI-_P_OMEGA_N}}
\end{subfigure}%
\begin{subfigure}{.5\textwidth}
\centering
\includegraphics[width=\textwidth]{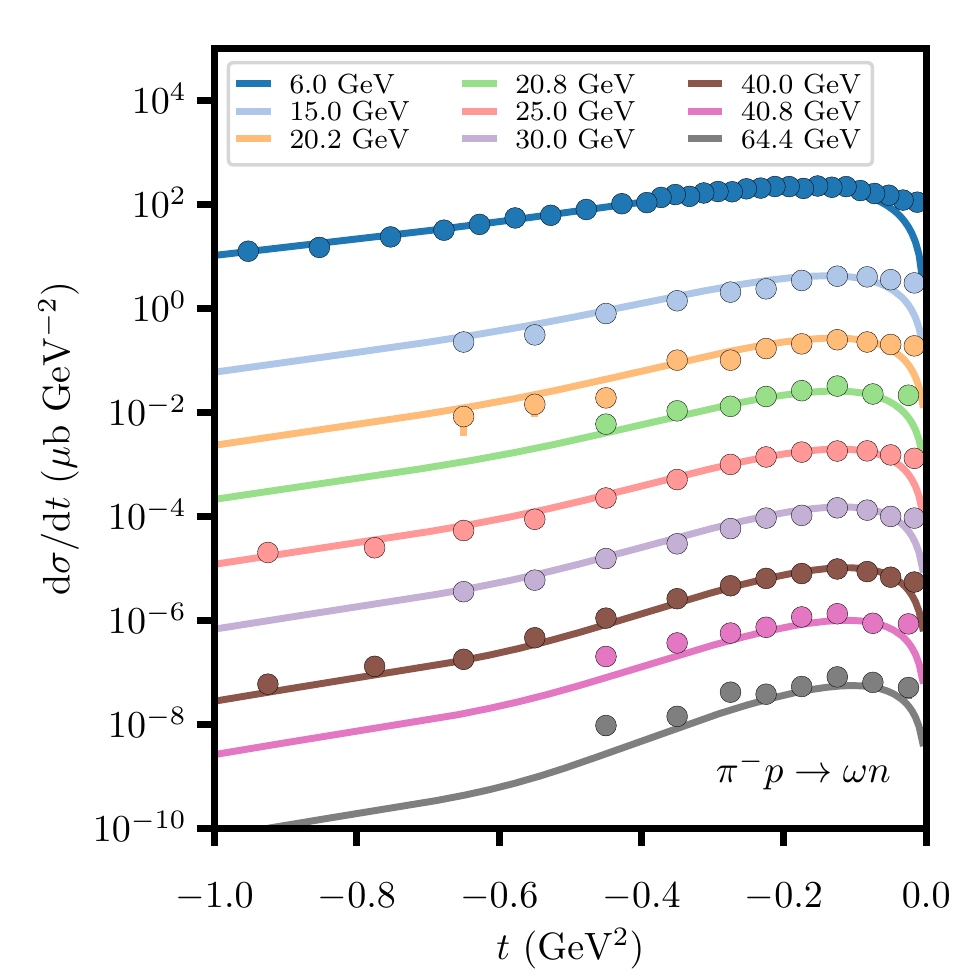}
  \caption{$\pi^- p \to \omega n$, $\plab < 100\gev$\label{fig:PI-_P_OMEGA_N_full}}
\end{subfigure}
\caption{Results for the CEX $\omega$ production channels. Scaling and conventions are  as in Fig.~\ref{fig:PI_PI_RHO}.\label{fig:OMEGAN}}
\end{figure*}

The $\Gamma$ function in Eq.~\eqref{eq:NWSZ_beta} introduces zeros in the cross sections for reactions dominated by the $\rho$ exchange, \eg $\pi^- p \to \pi^0 n$. In this case, daughter poles or other subleading singularities become relevant as they tend to fill in the zeros~\cite{Irving:1977ea,Szczepaniak:2014bsa}. In the SU(3)-EXD we focus only on the leading natural Regge poles and do not include any subleading Regge contribution. Instead of introducing new amplitude components to fill in the dips, we prefer to work with the well defined Regge pole model. Therefore, we must reduce the contribution from dip regions in
$\rho$-only dominated channels in Figs.~\ref{fig:PI_PI_RHO} and~\ref{fig:PI-_P_OMEGA_N}, by rescaling the data error bars using a factor $f(t) = 1/\abs{t-t_0}$, where $t_0$ is defined such that $\alpha_\rho(t_0) = 0$. Such a rescaling is required in order for the fit to be less sensitive to kinematic regions dominated by components beyond the leading pole model. The goal of this work is to detect the regions where the Regge pole approximation holds best, by manually varying the kinematic regions. Additionally, the available data is subject to sizable and uncontrolled systematic errors. Therefore, we aim to provide a qualitative description, and do not report the $\chi^2$ values and errors of the fits.

The global SU(3)-EXD fit contains nine free parameters, which are given in  Table~\ref{tab:reduced_couplings_SU3}. The other three are  the $\eta-\eta'$ mixing angle $\theta_P$ and the single- and non-helicity-flip exponential factors $b_{sf}$ and $b_{nf}$, respectively. In total, $N_\text{data}=1271$ high-energy forward scattering data points are fitted, with $\plab \geq 5\gev$ and $0 \leq -t \leq 0.8\gev^2$. For the total cross sections, we consider data at slightly higher energies,  $\plab \geq 10\gev$, since the Regge-based model in~\cite{pdg} matches best above this energy. The parameters of the natural Regge trajectories are fixed to the values given in 
 Table~\ref{tab:trajectories}.
Since there are too many amplitudes describing the production 
 of spin $3/2$ baryons compared to the available data, additional constraints are needed. 
 Specifically, we keep only a single term in the interaction Lagrangian of vector-meson--octet-baryon--decuplet-baryon couplings (VBD), \ie we set $g^{(2)}_{VBD} = g^{(3)}_{VBD} = 0$ in Eq.~\eqref{eq:L_VBD} for all VBD combinations. More details are given in  Appendix~\ref{sec:lagrangians}. Note that this sets all double flip components to zero, in agreement with the data. Additionally, it is worth mentioning that the reduced residue of the non-flip component is proportional to $t$ in the single-meson form in Table~\ref{tab:s_channel_residues}. Even though this is not required by factorization and angular-momentum conservation, the contribution is therefore required to vanish at $t=0$ in the single-meson exchange approximation. 
  
\begin{table}
\caption{Considered exchanges with their relevant quantum numbers and corresponding trajectories used as input for the global SU(3)-EXD fit ($\alpha(1)$) and as obtained from the unconstrained fit ($\alpha(2)$). The Mandelstam variable $t$ must be expressed in units $\gev^2$.\label{tab:trajectories}}
\begin{tabular}{|c|c|c|c|c|}
\hline
Reggeon & $I^{G \zeta \eta }$ & $l_e$ & $\alpha$(1) & $\alpha$(2)  \\
\hline \hline
$\rho$ & $1^{+-+}$ & $1$ & $0.5\phantom{0}+0.9\, t$ & $0.51 + 0.82\, t$ \\
$a_2$ & $1^{-++}$ & $1$ & $0.5\phantom{0}+0.9\, t$ & $0.42+0.90\, t$ \\
$K^*$ & $1^{.-+}$ & $1$ & $0.35+0.9\, t$ & $0.35+0.9\, t$ \\
$K^*_2$ & $1^{.++}$ & $1$ & $0.35+0.9\, t$ & $0.35+0.9\, t$ \\
$b_1$ & $1^{+--}$ & $0$ & --- & $0.7\,t$\\
\hline
\end{tabular}
\end{table}

\begin{table}[tb]
\caption{Available differential cross section data and references. \label{tab:data_refs}}
\begin{tabular}{|L|l|}
\hline
\mbox{Reaction} & Data references\\
\hline\hline
\pi^+ p \to X &~\cite{pdg} \\
K^\pm N \to X	&~\cite{pdg} \\
\hline
\pi^- p \to \pi^0 n &~\cite{Barnes:1976ek},~\cite{Stirling:1965zz} \\
\pi^+ p \to \pi^0 \Delta^{++} &~\cite{Bloodworth:1974ji},~\cite{Honecker:1977me},~\cite{Schotanus:1970xf},~\cite{Scharenguivel:1971ev} \\
\hline
\pi^- p \to \eta n &~\cite{Apel:1978dv},~\cite{Bolotov:1974wt},~\cite{Dahl:1976em},~\cite{Daum:1980ve} \\
\pi^- p \to \eta' n &~\cite{Daum:1980ve},~\cite{Apel:1979ic}  \\
\pi^+ p \to \eta \Delta^{++} &~\cite{Bloodworth:1974ji},~\cite{Honecker:1977me} \\ 
\hline
K^+ n \to K^0 p &~\cite{Diebold:1974cx},~\cite{Gilchriese:1977ee},~\cite{Phelan:1975ay} \\
K^- p \to \overline{K}^0 n & ~\cite{Diebold:1974cx},~\cite{Gilchriese:1977ee},~\cite{Phelan:1975ay},~\cite{Ambats:1973hv},~\cite{Astbury:1965rra},~\cite{Binon:1981ha},~\cite{Bolotov:1974pr},~\cite{Brandenburg:1976yj}, \\
& ~\cite{Chaurand:1976sb},~\cite{Gallivan:1976cf}\\
K^+ p \to K^0 \Delta^{++} &~\cite{Gilchriese:1977ee},~\cite{Phelan:1975ay},~\cite{Brandenburg:1976yj},~\cite{Carney:1976kw},~\cite{Foley:1974ex} \\
K^- n \to \overline{K}^0 \Delta^- &~\cite{Gilchriese:1977ee},~\cite{Phelan:1975ay} \\
K^- p \to \overline{K}^0 \Delta^0 &~\cite{Brandenburg:1976yj},~\cite{Gallivan:1976cf} \\ 
\hline
K^- p \to \pi^- \Sigma^{*+} &~\cite{Chaurand:1976sb},~\cite{Baker:1979kf},~\cite{Ballam:1978rx},~\cite{Baubillier:1984pj} \\
K^- p \to \pi^- \Sigma^+ &~\cite{Chaurand:1976sb},~\cite{Baker:1979kf},~\cite{Ballam:1978rx},~\cite{Berglund:1978sm} \\
K^- p \to \pi^0 \Lambda &~\cite{Chaurand:1976sb},~\cite{AlHarran:1980br}\\
\pi^+ p \to K^+ \Sigma^{*+} &~\cite{Baker:1979kf},~\cite{Ballam:1978rx},~\cite{Bashian:1972cs} \\ 
\pi^+ p \to K^+ \Sigma^+ &~\cite{Baker:1979kf},~\cite{Ballam:1978rx},~\cite{Berglund:1978sm},~\cite{Bashian:1972cs},~\cite{Bitsadze:1984xq} \\
\pi^- p \to K^0 \Lambda &~\cite{Crennell:1972km},~\cite{Foley:1973ve},~\cite{Ward:1973ce} \\ 
\pi^- p \to K^0 \Sigma^0 &~\cite{Crennell:1972km},~\cite{Foley:1973ve},~\cite{Ward:1973ce} \\
K^- p \to \eta \Lambda	&~\cite{AlHarran:1980br,Chaurand:1976sb} \\
K^- p \to \eta' \Lambda	&~\cite{AlHarran:1980br,Chaurand:1976sb} \\
 \hline
\pi^- p \to \omega n &~\cite{Apel:1979yx},~\cite{Dahl:1976ky},~\cite{Shaevitz:1975ir} \\
\pi^+ n \to \omega p &~\cite{Paler:1972fm} \\
\hline
\end{tabular}
\end{table}
\begin{table}
\caption{Reduced SU(3) couplings obtained from a global SU(3)-EXD fit. Fixed couplings are indicated by an asterisk.\label{tab:reduced_couplings_SU3}}
\begin{tabular}{|C|C|C|}
\hline
\text{Top vertices} & \text{Bottom vertices} & \text{Damping } (\gev^{-2})\\
\hline\hline
\theta_P = -0.14	& g^{v,D}_{VBB} = -1.29 & b_\text{sf} = 0.54 \\
g_{VPP} = -4.2^*		& g^{v,F}_{VBB} = 2.35 & b_\text{nf} = 1.31 \\
g_{VVP} = 45.25\gev^{-2}					& g^{t,D}_{VBB} = 6.93 &  \\
                    & g^{t,F}_{VBB} = 3.64 &\\
                    & g^{(1)}_{VBD} = -7.11 &\\
\hline
\end{tabular}
\end{table}

\begin{figure*}[h]
\centering
\includegraphics[width=0.5\textwidth]{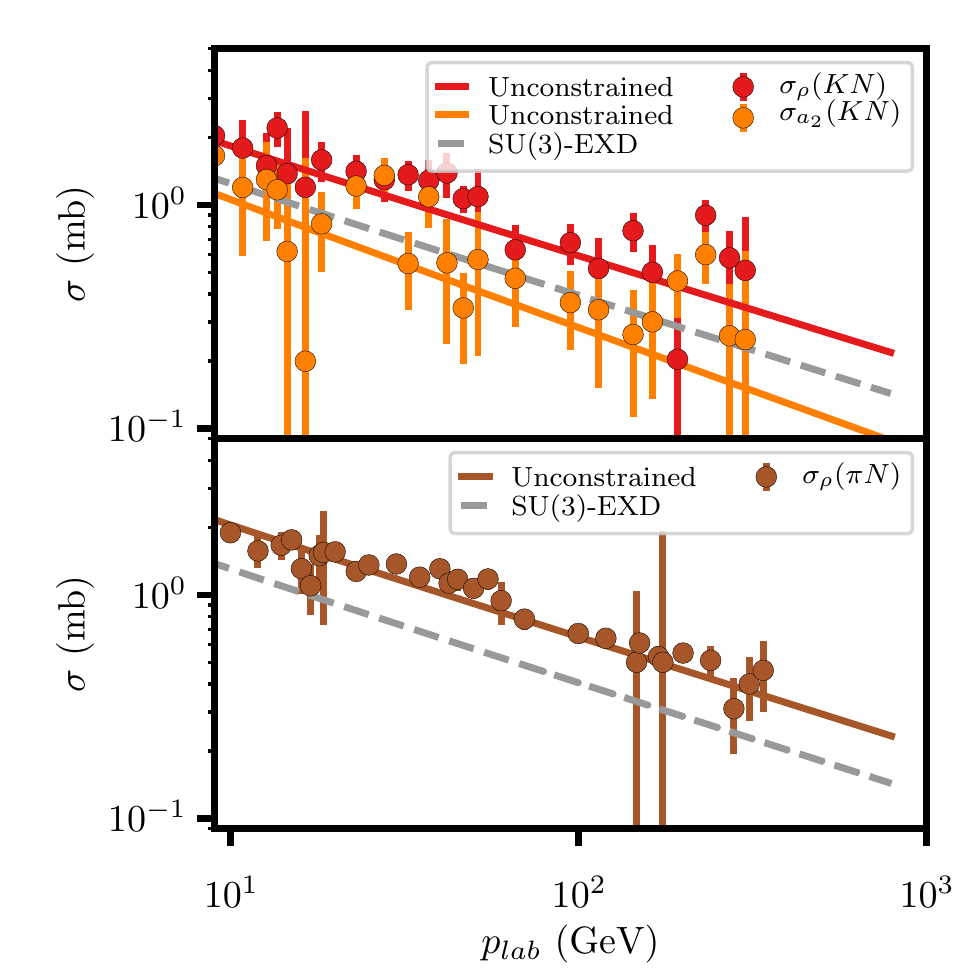}
\caption{Total cross section combinations in Eqs.~\eqref{eq:TOTCS_comb1}-\eqref{eq:TOTCS_comb3}. The dashed and solid lines represent the global SU(3)-EXD and isospin constrained fit respectively. \label{fig:TOTCS}}
\end{figure*}

\begin{figure*}[h]
\begin{subfigure}{.5\textwidth}
\centering
\includegraphics[width=\textwidth]{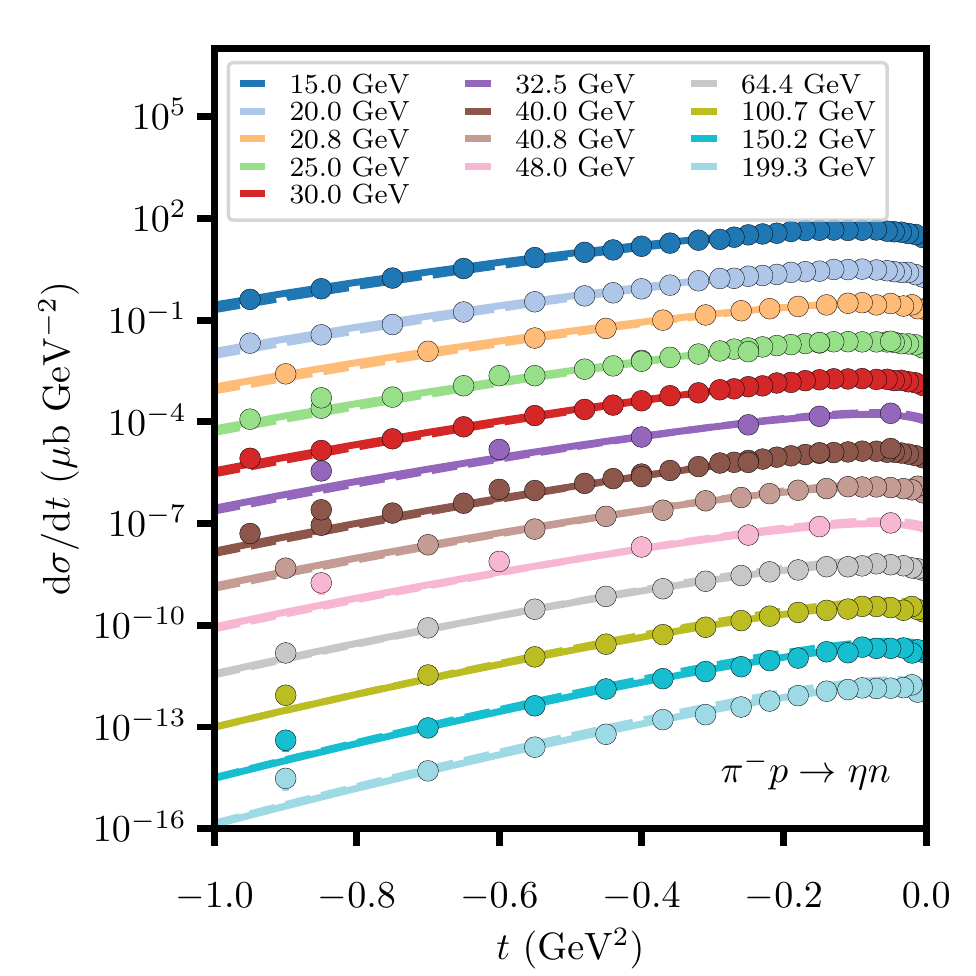}
  \caption{$\pi^- p \to \eta n$\label{fig:PI-_P_ETA_N}}
\end{subfigure}%
\begin{subfigure}{.5\textwidth}
\centering
\includegraphics[width=\textwidth]{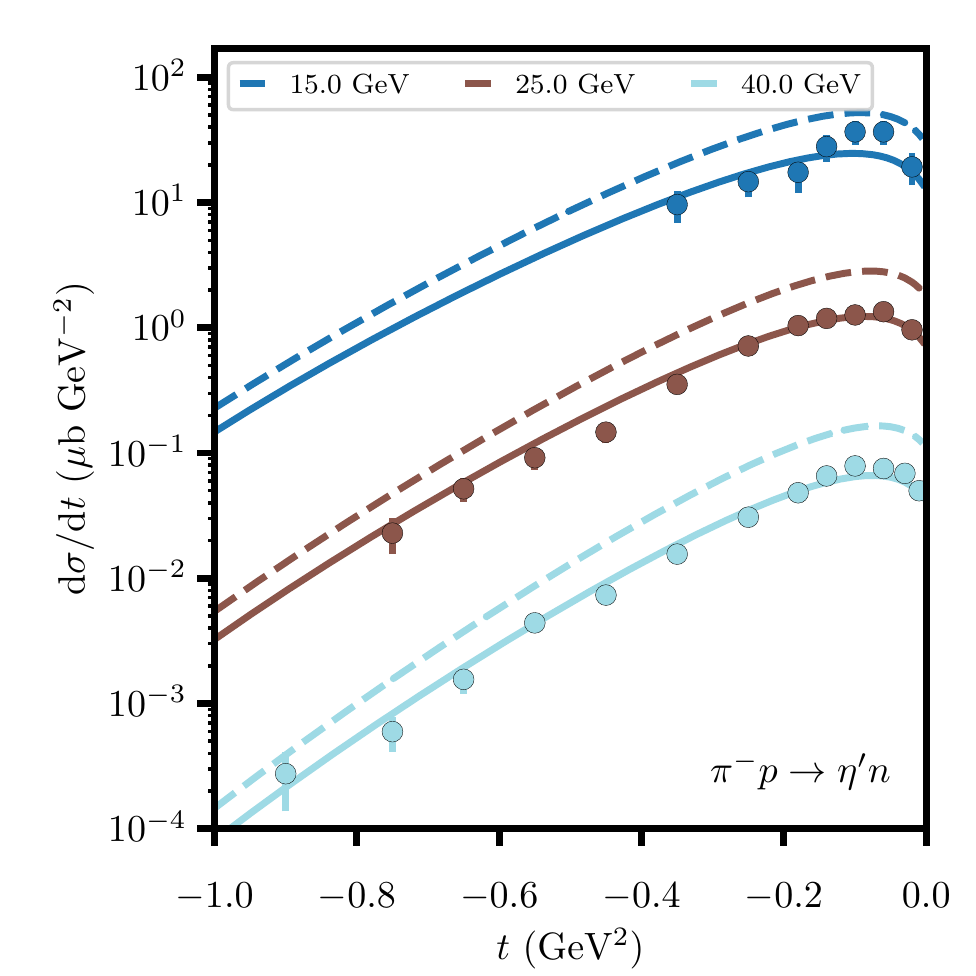}
  \caption{$\pi^- p \to \eta' n$\label{fig:PI-_P_ETAPRIME_N}}
\end{subfigure}%

\begin{subfigure}{.5\textwidth}
\centering
\includegraphics[width=\textwidth]{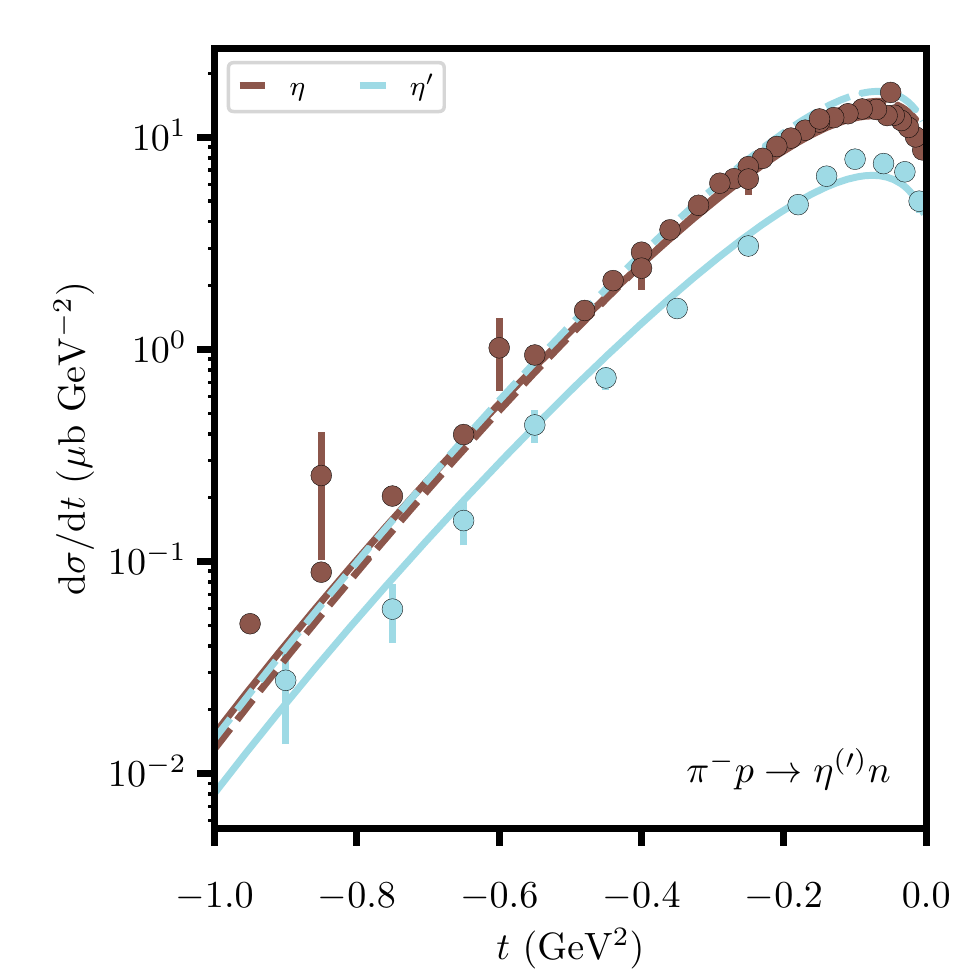}
  \caption{$\pi^- p \to \eta^{(\prime)} n$\label{fig:PI-_P_ETA_PRIME_N_overlap}}
\end{subfigure}%
\caption{Differential cross section for the channels dominated by $a_2$ exchange only. Scaling and conventions are as in Fig.~\ref{fig:PI_PI_RHO}. Figure~\ref{fig:PI-_P_ETA_PRIME_N_overlap} shows $\eta^{(\prime)} n$ for $\plab = 40\gev$.\label{fig:PI_ETAP_A2}}
\end{figure*}

\begin{figure*}[h]
\centering
\begin{subfigure}{.5\textwidth}
\centering
\includegraphics[width=\textwidth]{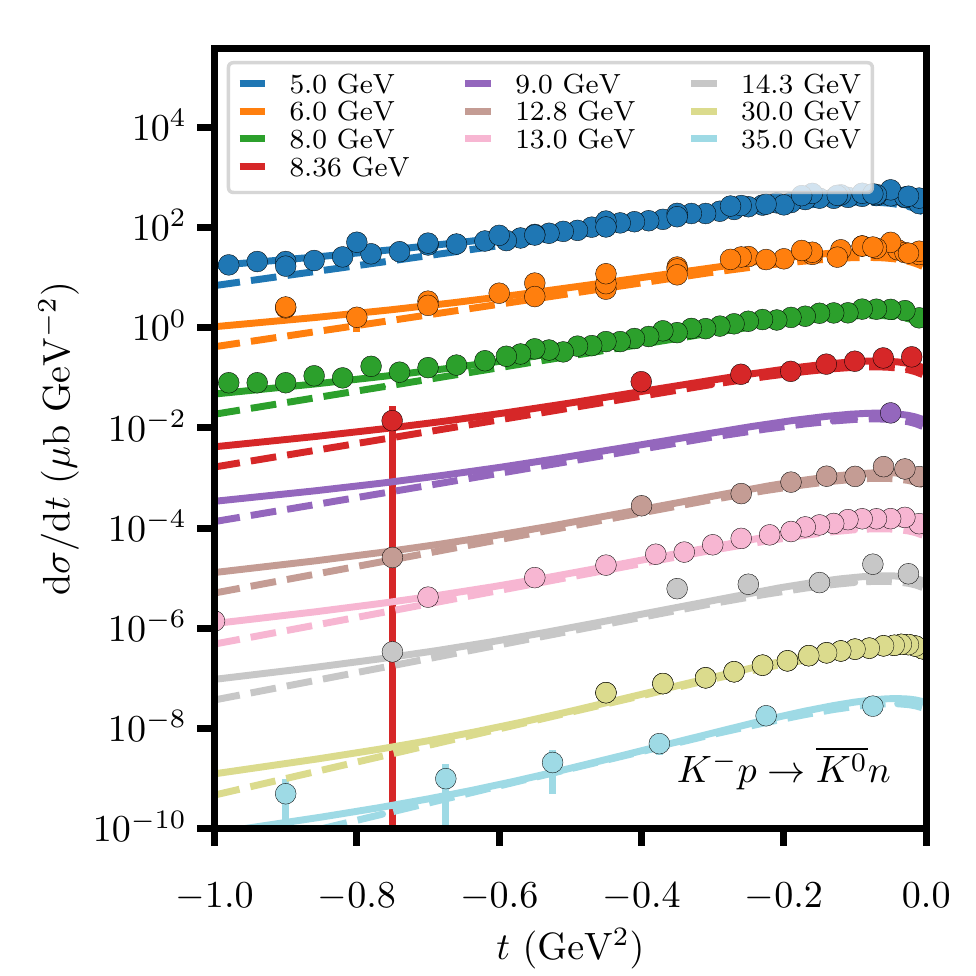}
  \caption{$K^- p \to \kbarz n$\label{fig:K-_P_KBAR0_N}}
\end{subfigure}%
\begin{subfigure}{.5\textwidth}
\centering
\includegraphics[width=\textwidth]{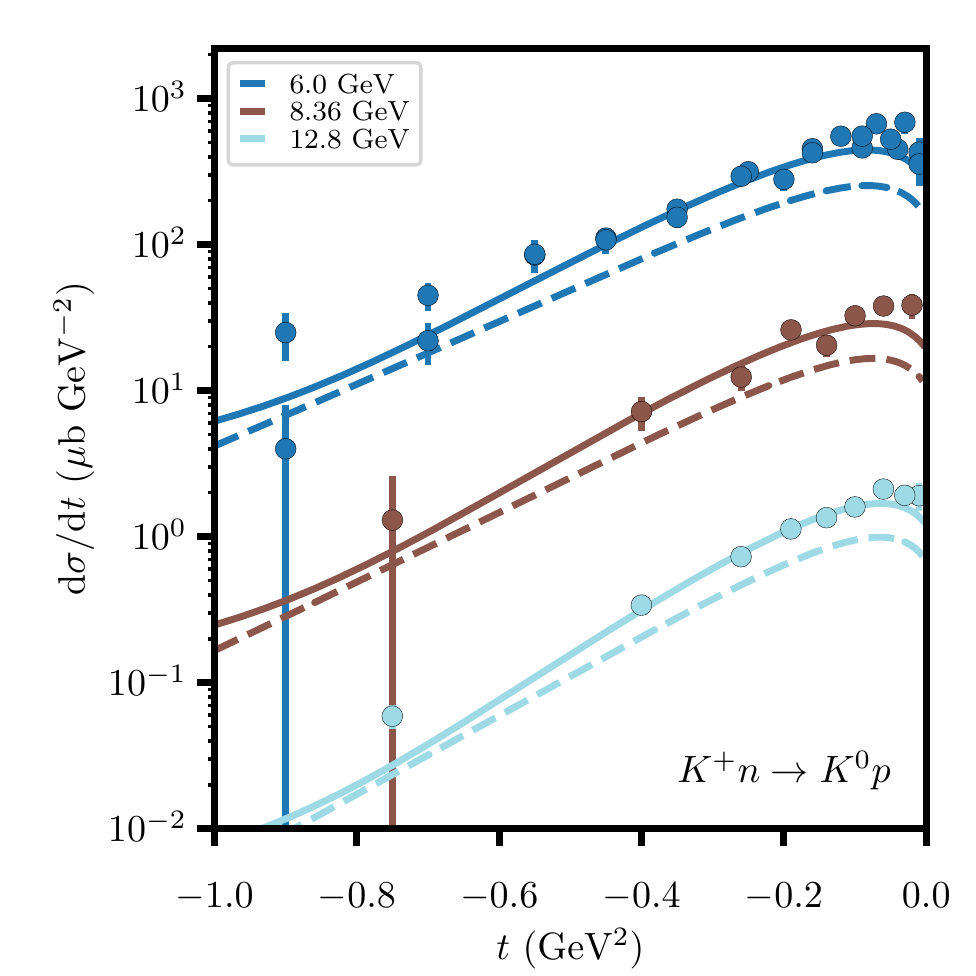}
  \caption{$K^+ n \to K^0 p$\label{fig:K+_N_K0_P}}
\end{subfigure}%
\caption{$KN\to KN$ CEX reaction data. Scaling and conventions are as in Fig.~\ref{fig:PI_PI_RHO}. \label{fig:KN}}
\end{figure*}

\begin{figure*}[h]
\centering
\begin{subfigure}{.5\textwidth}
\centering
\includegraphics[width=\textwidth]{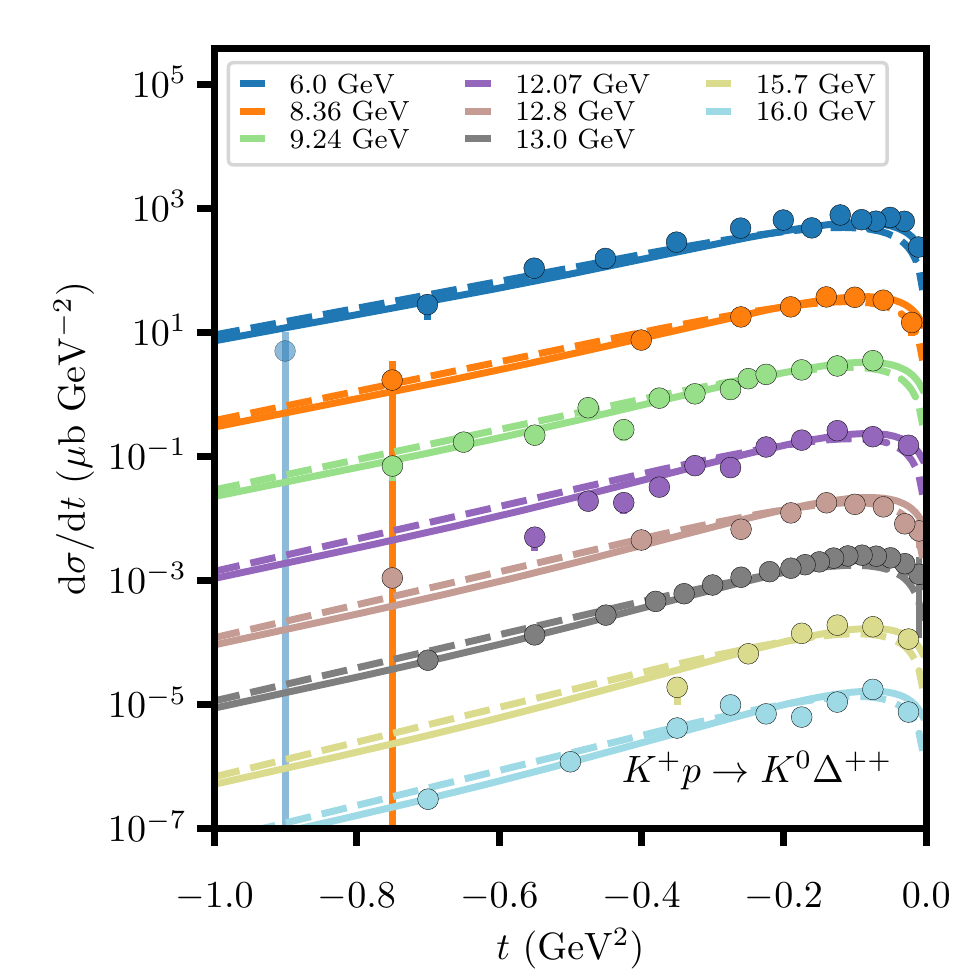}
  \caption{$K^+ p \to K^0 \Delta^{++}$\label{fig:K+_P_K0_DELTA_1232P33_++}}
\end{subfigure}%
\begin{subfigure}{.5\textwidth}
\centering
\includegraphics[width=\textwidth]{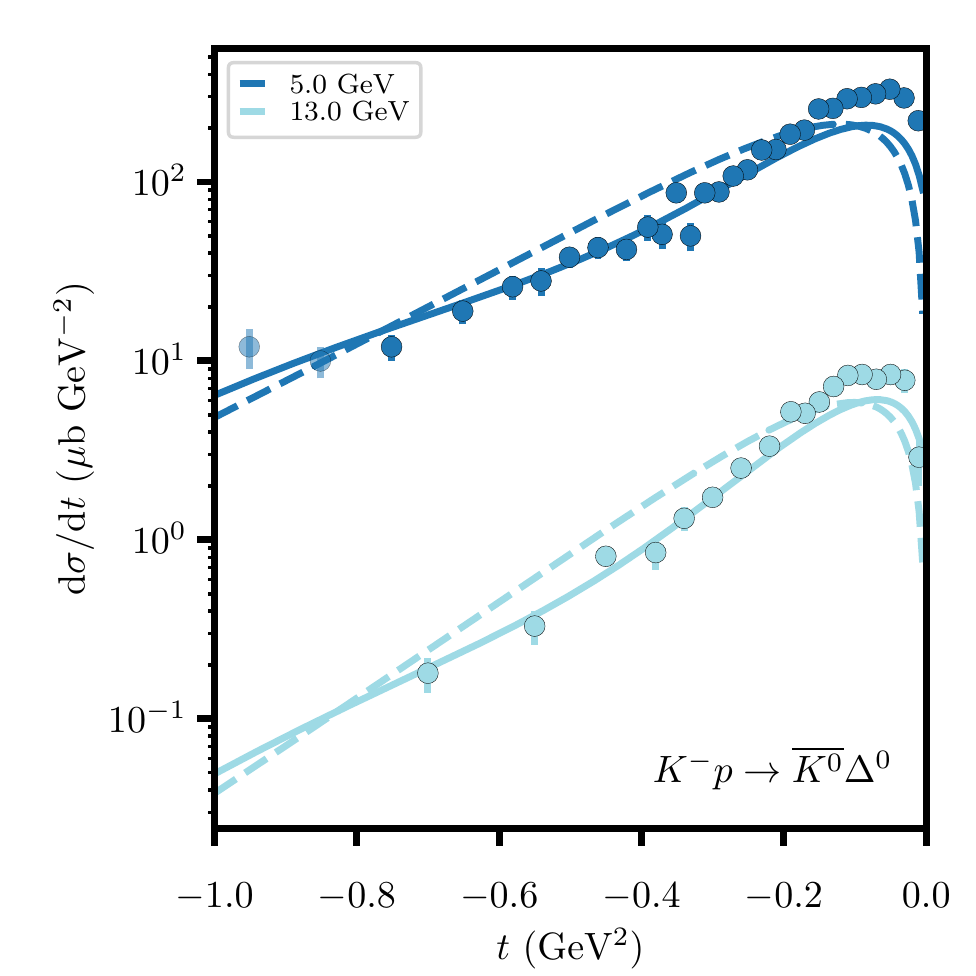}
  \caption{$K^- p \to \kbarz \Delta^0$\label{fig:K-_P_KBAR0_DELTA_1232P33_0}}
\end{subfigure}%
\caption{$KN\to K\Delta$ CEX reaction data. Scaling and conventions are  as in Fig.~\ref{fig:PI_PI_RHO}.\label{fig:KDELTA}}
\end{figure*}

\begin{figure*}[h]
\centering
\includegraphics[width=0.5\textwidth]{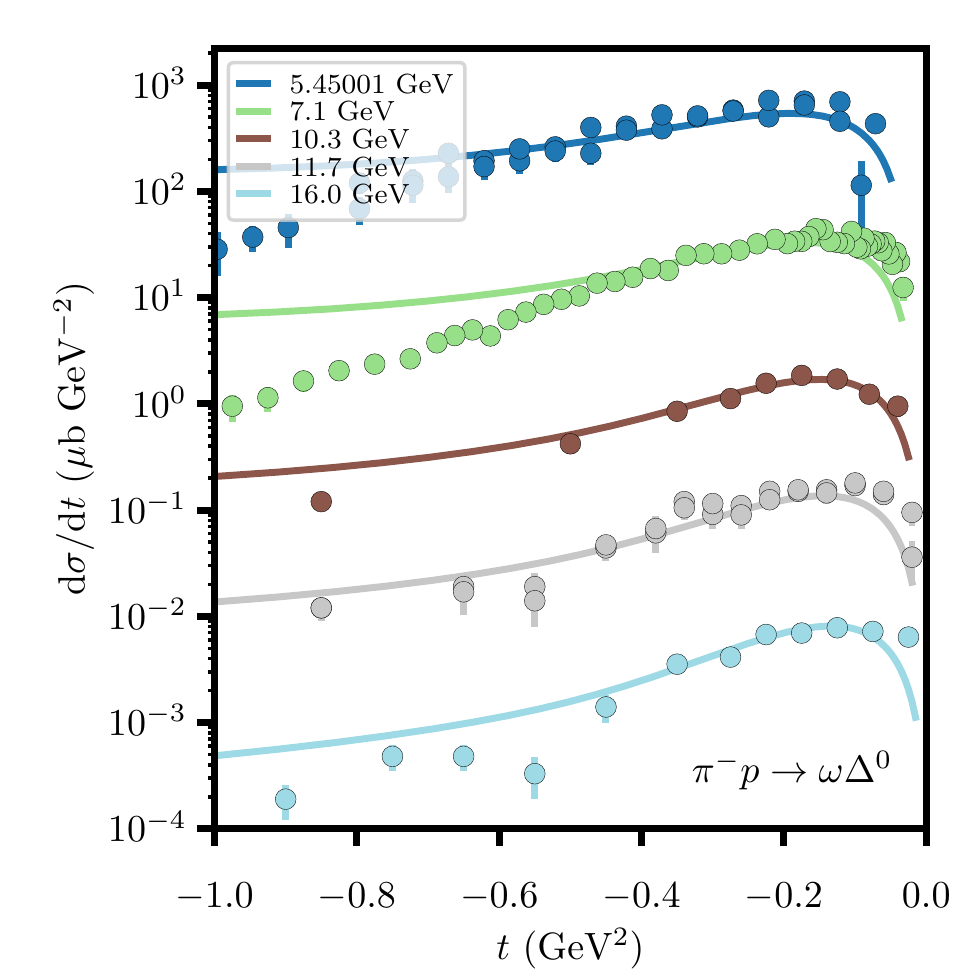}
\caption{Results for the CEX $\omega$ production channel $\pi^- p \to \omega \Delta^0$. Scaling and conventions are  as in Fig.~\ref{fig:PI_PI_RHO}.\label{fig:OMEGADELTA}}
\end{figure*}

\begin{figure*}[h]
\begin{subfigure}{.5\textwidth}
\centering
\includegraphics[width=\textwidth]{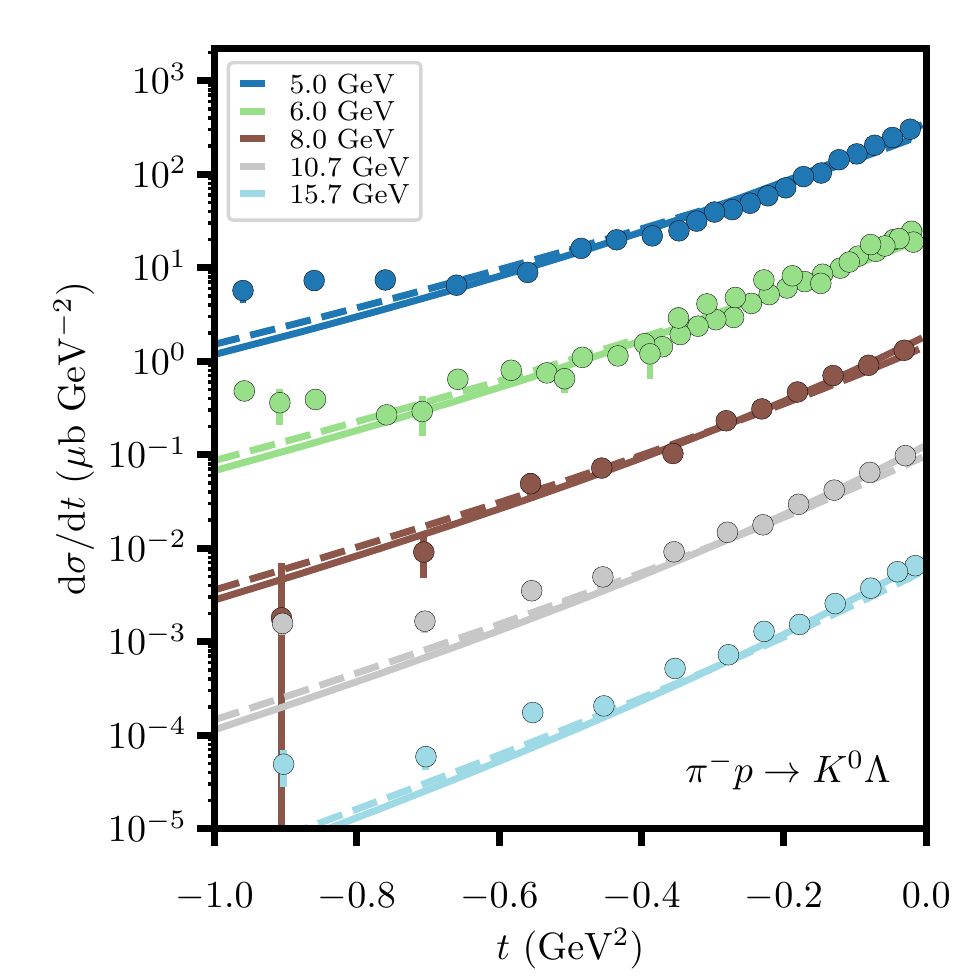}
  \caption{$\pi^- p \to K^0 \Lambda$\label{fig:PI-_P_K0_LAMBDA}}
\end{subfigure}%
\begin{subfigure}{.5\textwidth}
\centering
\includegraphics[width=\textwidth]{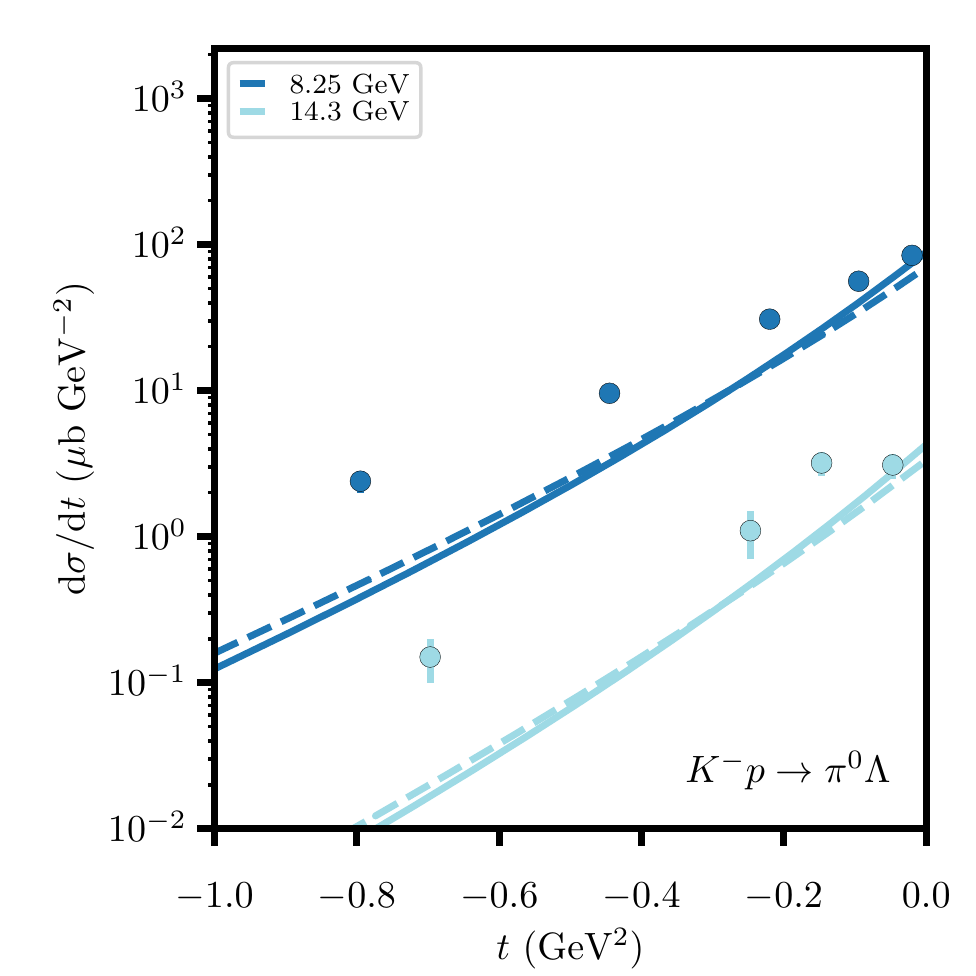}
  \caption{$K^- p \to \pi^0 \Lambda$\label{fig:K-_P_PI0_LAMBDA}}
\end{subfigure}%
\caption{Results for the strangeness exchange $\pi \Lambda$ and $K\Lambda$ production channels. Scaling and conventions are  as in Fig.~\ref{fig:PI_PI_RHO}.\label{fig:PIKLAMBDA}}
\end{figure*}

\begin{figure*}[h]
\begin{subfigure}{.5\textwidth}
\centering
\includegraphics[width=\textwidth]{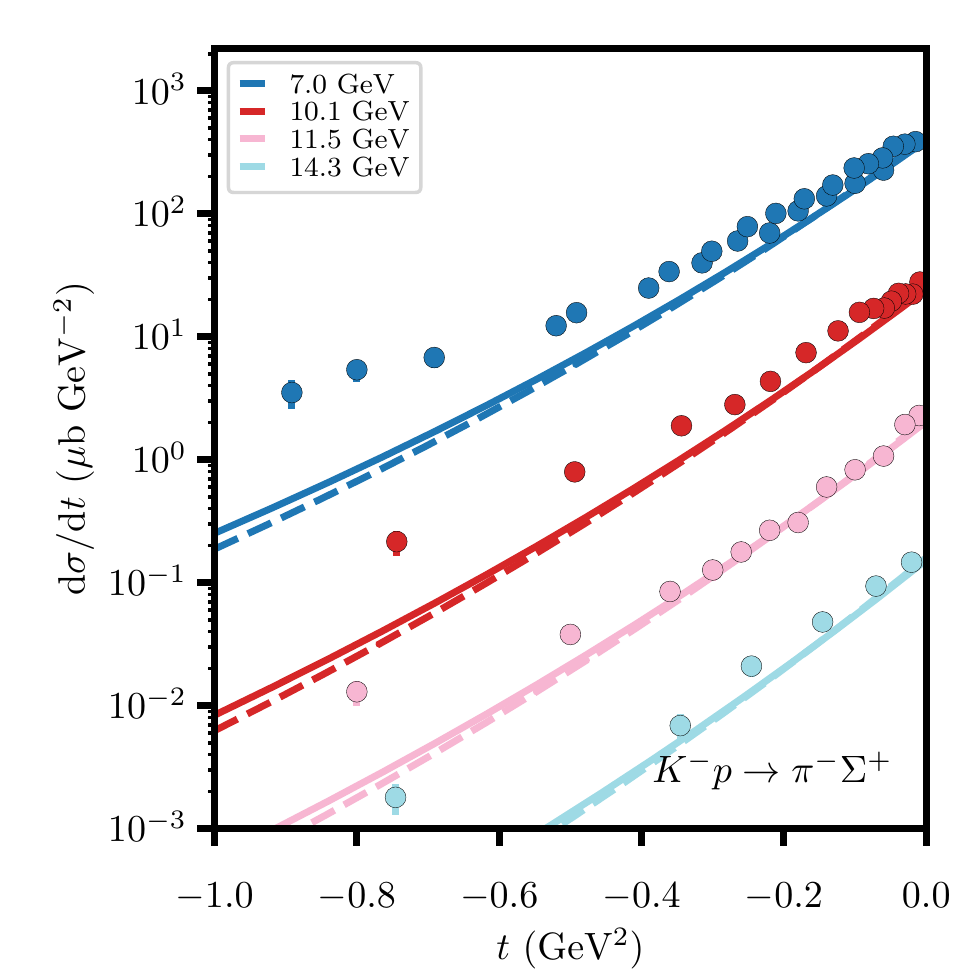}
  \caption{$K^- p \to \pi^- \Sigma^+$\label{fig:K-_P_PI-_SIGMA+}}
\end{subfigure}%
\begin{subfigure}{.5\textwidth}
\centering
\includegraphics[width=\textwidth]{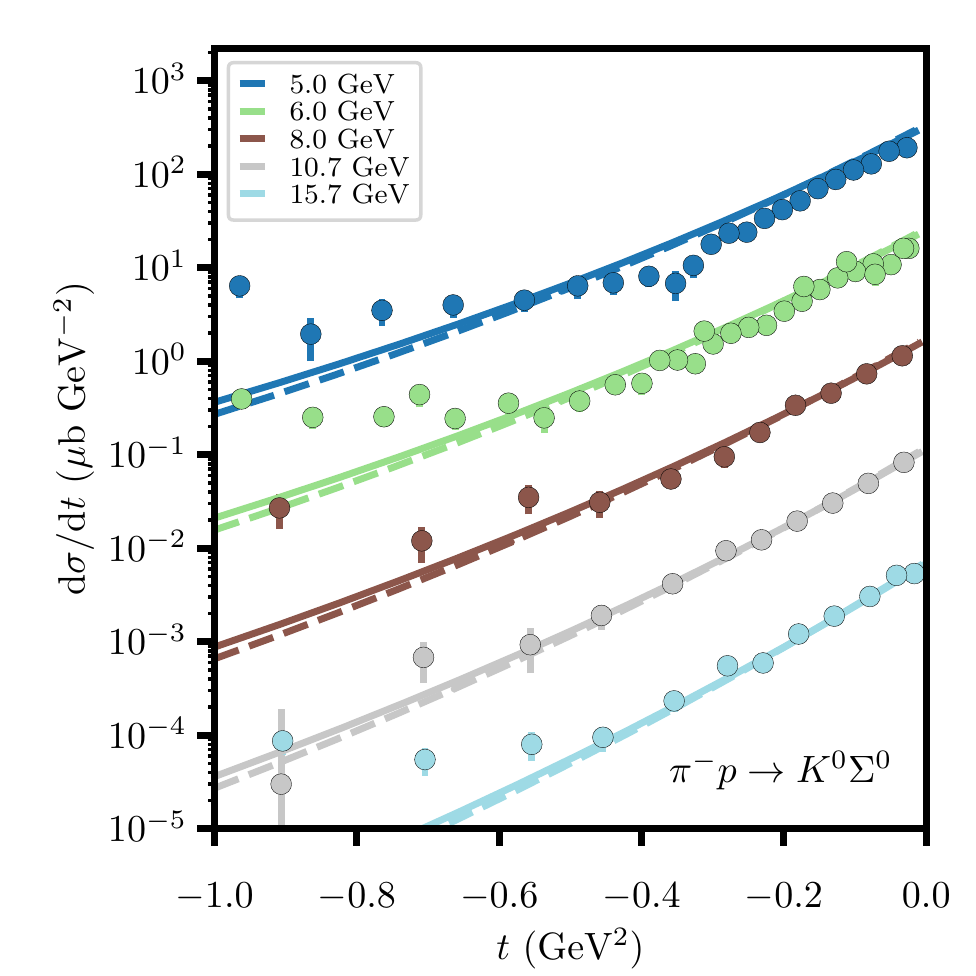}
  \caption{$\pi^- p \to K^0 \Sigma^0$\label{fig:PI-_P_K0_SIGMA0}}
\end{subfigure}
\begin{subfigure}{.5\textwidth}
\centering
\includegraphics[width=\textwidth]{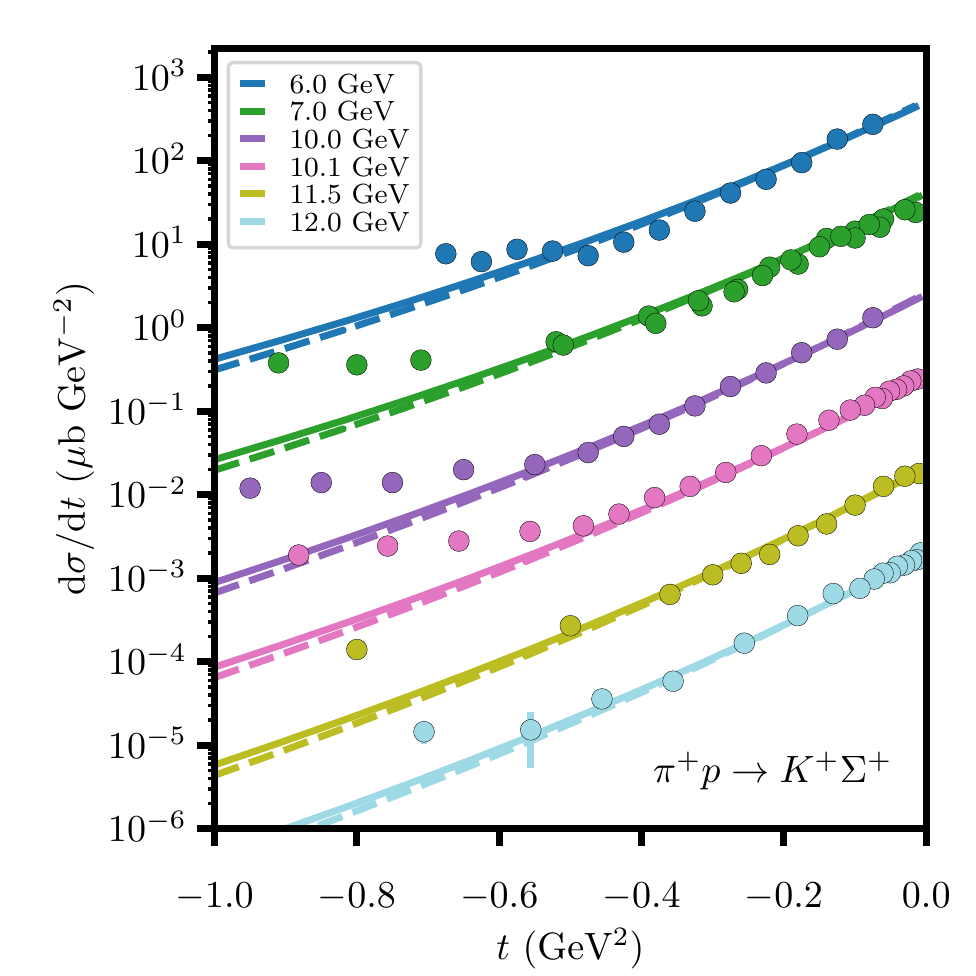}
  \caption{$\pi^+ p \to K^+ \Sigma^+$\label{fig:PI+_P_K+_SIGMA+}}
\end{subfigure}%
\caption{Results for the strangeness exchange $\pi \Sigma$ and $K\Sigma$ production channels. Scaling and conventions are  as in Fig.~\ref{fig:PI_PI_RHO}.\label{fig:PIKSIGMA}}
\end{figure*}

The comparison between the data and the model is shown in Figs.~\ref{fig:TOTCS}-\ref{fig:PIKSIGMASTAR}. The global SU(3)-EXD fit is stable and provides a remarkably good description of all the key features in the data. In Table~\ref{tab:SU3_residues_pole} we list all exchanges that contribute to the reactions we have analyzed and values of the residues derived from the fit. We do not take into account the exponential factor in Eq.~\eqref{eq:general_damping} when we extrapolate the residues to the pole, as it is expected to be a fair approximation in the physical region only. 
This will allow us to directly relate our extracted couplings to those in modern literature as discussed in Appendix~\ref{sec:couplings_estimations}. 
 The residues are computed from the couplings in Table~\ref{tab:reduced_couplings_SU3}. 

In the following, we discuss in detail the model predictions for the various channels. We consider first the appropriate combinations of total cross sections which are sensitive to $\rho$ and $a_2$ exchanges. 
The optical theorem relates the total cross section to the elastic amplitude at $t=0$ via
\begin{align}
\sigma(1 + 2 \to X) = \frac{\sum_{\mu_1 \mu_2} \Im A_{\mu_2 \mu_1 \mu_2 \mu_1} (s,t = 0)}{(2s_1 + 1)(2s_2 + 1) S_{12}(s)} \,,\label{eq:totcs_expression}
\end{align}
where $S_{ij}$ is a kinematic function defined in Appendix~\ref{sec:conventions}. The contribution from the individual Regge poles to the elastic amplitudes in Eq.~\eqref{eq:totcs_expression} is given by
\begin{subequations}
\begin{align}
A(\pi^\pm p) &= \pomeron + f_2 \mp \rho\,, \\
A(K^\pm p) &= \pomeron + f_2 \mp \rho + a_2 \pm \omega\,, \\
A(K^\pm n) &= \pomeron + f_2 \pm \rho - a_2 \pm \omega \,.
\end{align}
\end{subequations}
Hence, defining the following linear combinations,
\begin{subequations}
\begin{align}
\sigma_\rho(KN) =& \phantom{-} \sigma(K^- p) - \sigma(K^- n) \nonumber \\
&- \sigma(K^+ p) + \sigma(K^+ n) \,,\label{eq:TOTCS_comb1}\\
\sigma_{a_2}(KN) =&  \phantom{+} \sigma(K^- p) - \sigma(K^- n) \nonumber \\
&+ \sigma(K^+ p) - \sigma(K^+ n)\,, \\
\sigma_\rho(\pi N) =& \phantom{+}  \sigma(\pi^- p) - \sigma(\pi^+ p)\,,\label{eq:TOTCS_comb3}
\end{align}
\end{subequations}
one can investigate the individual contributions from the $\rho$ and $a_2$ exchanges. 
 In the case of exact SU(3) and EXD, all three cross section combinations must be equal. Inspecting the results shown in Fig.~\ref{fig:TOTCS}, one sees that $\sigma_{\rho}(KN)$ and $\sigma_{a_2}(KN)$ differ by a few mb only, indicating a small violation of exchange degeneracy. The compatibility of $\sigma_{\rho}(KN)$ and $\sigma_{\rho}(\pi N)$ illustrates that the SU(3) symmetry is well respected in $KN$ and $\pi N$ scattering.

The $\Gamma$ function in Eq.~\eqref{eq:regge_propagator} is introduced to remove spurious spin exchange components. 
The spin $0$ exchange term in the $a_2$ trajectory at $\alpha_{a_2}(t) = 0$ corresponds to an unphysical pole with negative mass squared. As discussed earlier, due to EXD, a zero in the residue for $a_2$ exchange forces a zero in the residue of the $\rho$ exchange, even though the Regge propagator for the $\rho$ does not contain a pole at $\alpha_\rho(t)=0$. The residue zero results in the vanishing of the cross section, which can be observed in all reactions where $\rho$ exchange dominates, {\it i.e.} in $\pi N \to \pi N$ and $\pi N \to \pi \Delta$ in Fig.~\ref{fig:PI_PI_RHO}.

The SU(3) predictions for $\eta^{(\prime)}$ production are depicted in Figs.~\ref{fig:PI-_P_ETA_N} and \ref{fig:PI-_P_ETAPRIME_N}, and an overlay of both reactions for $\plab = 40\gev$ is given in Fig.~\ref{fig:PI-_P_ETA_PRIME_N_overlap}. The comparison between $\eta$ and $\eta'$ fits can be used to extract information on the pseudoscalar mixing. Using SU(3) constraints, one estimates the relative couplings as (see Appendix~\ref{sec:lagrangians})
\begin{align}
\frac{g_{a_2 \pi \eta'}}{g_{a_2 \pi \eta}} = \frac{S_T \cos \theta_P + \sin \theta_P}{\cos \theta_P - S_T \sin \theta_P} \,.
\end{align}
As discussed in Appendix~\ref{sec:lagrangians}, the Okubo--Zweig--Iizuka (OZI) $s\bar s$ suppression rule requires that the relative coupling of the singlet $\eta_1$ and octet $\eta_8$ components is given by $S_T = \sqrt{2}$. 
Hence, for mixing angles $\theta_P \approx -0.17$ and under the OZI assumption ($S_T = \sqrt{2}$), one finds $g_{a_2 \pi \eta'} \approx g_{a_2 \pi \eta}$. The exact size of the pseudoscalar mixing angle $\theta_P$ is unknown, but the various theoretical estimates suggest values in the range $-0.38\lesssim \theta_P \lesssim -0.17$~\cite{Feldmann:1998vh,Goity:2002nn,Aubert:2006cy,Mathieu:2010ss,Escribano:2015nra,Osipov:2015lva}.

The fit results in a good correspondence to $KN \to KN$ and $KN \to K\Delta$ CEX in Figs.~\ref{fig:KN} and~\ref{fig:KDELTA}. In these reactions, both $\rho$ and $a_2$ contribute. While the $\rho$ dominates the very forward region, the $a_2$ exchange fills up the dip of the $\rho$ in the neighborhood of $\alpha_\rho(t)=0$. 

For $\omega N$ production at very high energies in Fig.~\ref{fig:PI-_P_OMEGA_N}, a fit with only $\rho$ exchange shows the correct $s$-dependence over a wide energy range. The data also clearly show a dipping behavior near $\alpha_\rho(t)=0$, as expected for a pure $\rho$ contribution. The data on $\omega$ production in Fig.~\ref{fig:OMEGAN} are dominated by $\rho$ exchange at very high energies $\plab \geq 100\gev$, since in the forward direction the $b_1$ exchange contribution is suppressed by a factor of $s^{-1/2}$ relative to the natural exchange. In the fits, we wish to determine the residues of only the leading Regge poles that are constrained by multiple channels. Therefore, we consider only $\plab \geq 100\gev$ to isolate  the $\rho$ component, and neglect the $b_1$ exchange. The data on $\omega \Delta$ production are rather scarce and at energies sensitive to the $b_1$ exchange (see Fig.~\ref{fig:OMEGADELTA}). Therefore, we do not consider them in the global SU(3)-CEX fit.

Finally we consider strangeness exchange Figs.~\ref{fig:PIKLAMBDA}--\ref{fig:PIKSIGMASTAR}. The effective trajectories\footnote{The effective trajectories are obtained by fitting $\alpha_\text{eff}(t)$
\begin{align}
\frac{\diffd \sigma}{\diffd t} = f(t) \left(\frac{s}{s_0}\right)^{2\alpha_\text{eff}(t) - 2} \,,\nonumber
\end{align}
to the $s$-dependence at fixed $t$, where $f(t)$ is a fitting parameter which may be different for all $t$.} obtained in~\cite{Foley:1973ve} from $\Lambda$ and $\Sigma^0$ production data, are much flatter than the ones compatible with $\alpha_{K^*}(t = m_{K^*}^2) = 1$ and $\alpha_{K^*}(t=m_{K^*_2}^2) = 2$ used in this work. They obtain $\alpha_\text{eff}^0 = 0.32$ and $\alpha_\text{eff}^1 = 0.23-0.43\gev^{-2}$. This disagreement indicates that secondary contributions (such as additional poles) are present at higher $-t$. The global fit indeed does not reproduce the high-$\abs{t}$ region. However, a very good agreement is found in the very forward region. This kinematic domain follows the $s$-dependence compatible with our trajectories in Table~\ref{tab:trajectories}.

For both $\Lambda$ and $\Sigma$ production, an inconsistency is observed for the time-reversal related reactions in Figs.~\ref{fig:PIKLAMBDA} and~\ref{fig:PIKSIGMA} respectively. The mismatch is related to the time-reversal symmetry  of the $\beta^{K^*_{(2)} \pi K}(t)$ vertex imposed in the fit~\cite{Irving:1977ea}. The latter requires $\beta^{K^*_{(2)} \pi K}(t) = -\beta^{K^*_{(2)} K \pi}(t)$, while the bottom vertex remains the same. Under the assumption of EXD, the cross section of the time-reversed reactions are therefore expected to be the same. 
The SU(3) relations force the $\Lambda$ (and to a lesser extent $\Sigma$) production to be dominated by helicity non-flip contributions in the forward direction\footnote{Care must be taken when interpreting the pole couplings in Table~\ref{tab:SU3_residues_pole}: since the $t$-dependence of the different helicity couplings differs, the relative size of the residues in the physical region might be quite different than pole values.}. Note that the non-flip contributions respect the EXD and SU(3) relations very well.

\section{Unconstrained fit}\label{sec:unconstrained_fit}
In the unconstrained fit, we relax the EXD and SU(3) constraint and only  keep SU(2), isospin, as a good symmetry.  
We also fit the Regge trajectory parameters of the $\rho$ and $a_2$ exchange.
In this fit, we can keep the single-particle approximation for the residues and fit the couplings in Table~\ref{tab:s_channel_residues} (without their SU(3) decomposition) for both vector and tensor exchanges. Indeed, when the EXD constraint is removed, the residues of the tensor exchanges must be determined independently. However, this approach is quite cumbersome. Additionally, this approach forces the residues to be restricted to the single-particle exchange residue. The tensor poles are quite far away from the physical region and it can no longer be expected that this approximation is reliable in the physical region. Therefore, in our next fit the $t$-dependence of the residues is no longer constrained by the single-particle model and we use  Eq.~\eqref{eq:residue_unconstrained_fit} instead. 
Hereby, the reduced residues are parametrized as 
\begin{align}\label{eq:residue_unconstrained_fit}
\betahat^{eif}_{\mu_i \mu_f}(t) = g^{eif}_{\mu_i \mu_f} e^{b_{\mu_i \mu_f}^{eif} t}\,,
\end{align}
where the constants $g^{eif}_{\mu_i \mu_f}$ and $b_{\mu_i \mu_f}^{eif}$ are all fitted independently, unless stated otherwise.

Abandoning the strict connection with the particle exchange model also implies, for example, that the non-flip component of the $VBD$ coupling is  no longer required  to vanish at $t=0$ (see $\betahat^{VBD}_{+\frac{1}{2} +\frac{1}{2}} \propto t$ for the single-particle residue in Table~\ref{tab:s_channel_residues}). All this significantly increases the number of parameters from 9 to 110, and we fit them in steps including a few reactions at the time.
The $t$-dependence in our fit is now entirely absorbed into the exponential factor.

Next, we describe the step-wise fitting process. We take advantage of the fact that a given exchange
is related to a limited set of reactions.
In the first step, we determine the trajectory intercepts, $\alpha_{\rho}^0$, $\alpha_{a_2}^0$, and the parameters 
of the residues,   
 $\betahat^{\rho N N}_{++}$, $\betahat^{a_2 N N}_{++}$, $\betahat^{\rho N N}_{-+}$, $\betahat^{a_2 N N}_{-+}$, $\betahat^{\rho K K}$, $\betahat^{a_2 K K}$, $\betahat^{\rho \pi \eta^{(\prime)}}$ and $\betahat^{a_2 \pi \eta^{(\prime)}}$
 {\it cf}. Eq.~\eqref{eq:general_damping}.
  The slopes $\alpha_\rho^1$ ($\alpha_{a_2}^1$) follow from the requirement that $\alpha_\rho(t=m_\rho^2) = 1$ ($\alpha_{a_2}(t=m_{a_2}^2) = 2$). 
  The intercepts are allowed to vary in the range $0.4 \leq \alpha_{(\rho,a_2)}^0 \leq 0.55$. The results of these fits are depicted in Figs.~\ref{fig:TOTCS},~\ref{fig:PI-_P_PI0_N},~\ref{fig:PI_ETAP_A2}, and~\ref{fig:KN} Assuming $S_T = \sqrt{2}$, we obtain the mixing angle $\theta_P = -0.33$. This angle is compatible with the values found in the recent literature~\cite{Escribano:2015nra,Osipov:2015lva}. 
   Inspecting the results in Fig.~\ref{fig:PI-_P_ETA_N} one finds that the cross section rises rapidly for $-t \geq 0.8\gev^2$ and cannot be described within the pure Regge-pole picture. Therefore, this kinematic domain is excluded in the channels that follow. 

\begin{figure*}[h]
\begin{subfigure}{.5\textwidth}
\centering
\includegraphics[width=\textwidth]{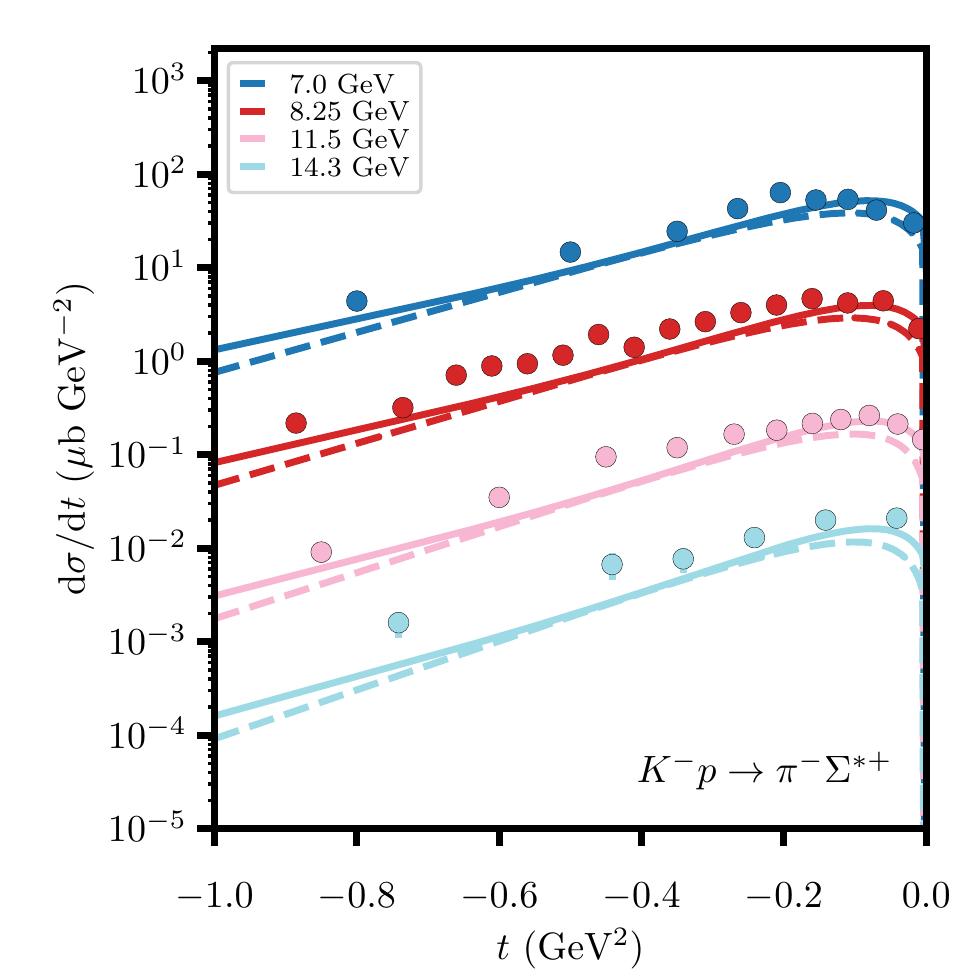}
  \caption{$K^- p \to \pi^- \Sigma^{*+}$\label{fig:K-_P_PI-_SIGMA_1385P13_+}}
\end{subfigure}%
\begin{subfigure}{.5\textwidth}
\centering
\includegraphics[width=\textwidth]{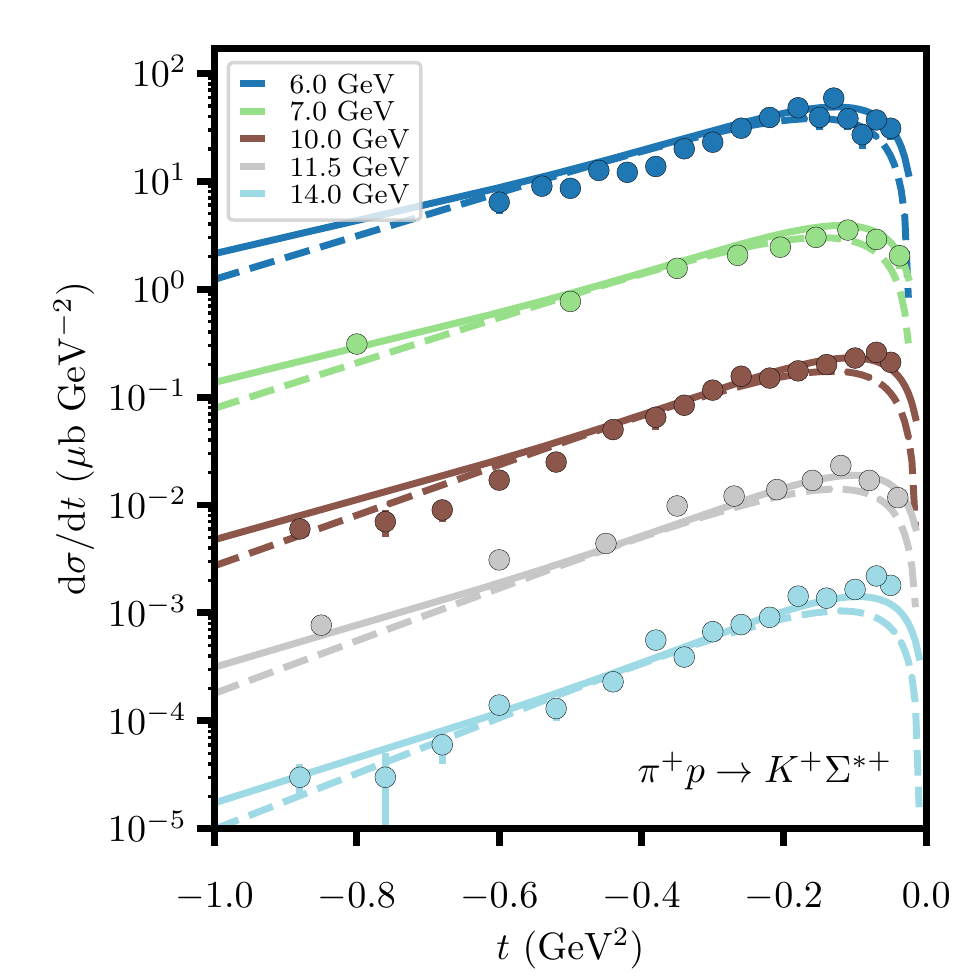}
  \caption{$\pi^+ p \to K^+ \Sigma^{*+}$\label{fig:PI+_P_K+_SIGMA_1385P13_+}}
\end{subfigure}%
\caption{Results for the strangeness exchange $\pi \Sigma^*$ and $K\Sigma^*$ production channels. Scaling and conventions are  as in Fig.~\ref{fig:PI_PI_RHO}.\label{fig:PIKSIGMASTAR}}
\end{figure*}

With the couplings at the top vertex determined through the residues listed above, we proceed to determine the bottom couplings, \ie  $\betahat^{\rho N \Delta}_{\mu_N \mu_\Delta}$  and $\betahat^{a_2 N \Delta}_{\mu_N \mu_\Delta}$, from a combined fit to  $\pi N \to \eta \Delta$, $\pi N \to \pi \Delta$ and $K N \to K\Delta$ cross sections. The large number of helicity couplings leads to a rather unconstrained fit and thus, based on the result of the  SU(3)-EXD fit, we eliminate the  double-flip components. Furthermore, we keep the ratio of the two single-flip components $g^{e N \Delta}_{+\frac{1}{2} +\frac{3}{2}}/g^{e N \Delta}_{-\frac{1}{2} +\frac{1}{2}}$ ($e = \rho,a_2$) fixed to the value obtained from the SU(3)-EXD fit. Their exponential $t$-dependence is assumed to be the same and is fitted to the data. The $t$-dependence of the non-flip components is fixed to the SU(3)-EXD values. The results of the fit are depicted in Figs.~\ref{fig:PI+_P_PI0_DELTA_1232P33_++}, and~\ref{fig:KDELTA}.
Relaxing the condition $\betahat^{\rho N \Delta}_{+\frac{1}{2} +\frac{1}{2}}(t = 0) = 0$ imposed by the single-particle exchange correspondence in Table~\ref{tab:s_channel_residues} seems to slightly improve the fit  at forward angles. This effect is even clearer for $\Sigma^{*}$ production channels in Fig.~\ref{fig:PIKSIGMASTAR}.

Using the $\betahat^{\rho N N}$ couplings extracted in the previous fitting steps, one can now determine the $\betahat^{\rho \pi \omega}$ residue from a fit to the $\omega N$ production data at very high energies. At forward angles and low energies the $b_1$ exchange does not represent the full strength required to reproduce the cross section of $\omega$ production. Notice that the NWSZ at $\alpha_{b_1}(t) = 0$ would force the cross section to vanish near $t\approx 0$. This mismatch is typically associated with the existence of  a trajectory with quantum numbers $I^{G\zeta \eta} = 1^{++-}$, with the lowest spin meson located on the trajectory being the yet undiscovered $\rho_2$ ($I^GJ^{PC} = 1^+2^{--}$)~\cite{Irving:1977ea,Irving:1974ak}\footnote{These quantum numbers are not exotic (only the $0^{--}$ is) and both the quark model and lattice QCD results predict the existence of such states~\cite{Godfrey:1985xj,Dudek:2013yja}. There are some experimental indications of the existence of $\rho_2$ and $\omega_2$ mesons~\cite{Anisovich:2002su,Anisovich:2011sva}. However, these states have been  observed  by  a  single  group and are poorly established, thus needing confirmation~\cite{pdg}.}  Due to its positive signature, this contribution is not required to vanish at $\alpha_{\rho_2}(t) = 0$, and might be even more important at forward angles than the $b_1$ exchange, provided that $\alpha_{\rho_2}(t)$ is similar to $\alpha_{b_1}(t)$; note however that $\alpha_{\rho_2}(t)$ is undetermined, as pointed out in~\cite{Nys:2016vjz}. This is because the NWSZ of the $b_1$ lies at $t=0$ which forces the $b_1$ contribution to vanish in the forward direction, independent of the factors in Eq.~\eqref{eq:sqrtt_factors}.
This lack of strength in our model in the forward direction hinders an unambiguous extraction of the $b_1$ couplings. While we do include a contribution from the $b_1$ exchange to absorb the different energy dependence at lower energies, we do not quote the results for the latter, since these couplings are unreliable. 

For the strangeness exchange channels the fits are somewhat more difficult  since one cannot separate the $K^*$ from $K^*_2$ exchanges due to the lack of definite $G$ parity. Additionally, there is less sensitivity to the trajectory parameters 
 due to a limited energy range in the data. 
  We therefore keep the trajectories fixed to the ones used in the global SU(3)-EXD fit. Additionally, the EXD constraint is imposed on the fits to reduce the number of free parameters. 
 In the global SU(3)-EXD fits shown in Figs.~\ref{fig:PI-_P_K0_LAMBDA}-\ref{fig:PI+_P_K+_SIGMA+}, and Table~\ref{tab:s_channel_residues}, it appears that the $0^- \frac{1}{2}^+$ production channels are dominated by Regge-pole non-flip contributions in the domain $0 \leq -t \leq 0.25\gev^2$. We can therefore carry out two-step fits, where we first determine the non-flip coupling from these very forward data. In a second step, we fix the non-flip coupling and extract the flip contribution from a full $t$ range fit. One observes that the helicity-flip couplings do not obey the SU(3) constraints well, in contrast to the non-flip contributions. Obtaining the helicity flip contributions from the fit turns out to be ambiguous. They depend strongly on the considered $t$ range and the final results deviate heavily from the SU(3)-EXD predictions. This issue was anticipated in the previous section, where we commented (based on the $s$-dependence) that the large $-t$ behavior of the cross section is dominated by contributions other than the Regge poles considered here.

\begin{figure*}[h]
\begin{subfigure}{.5\textwidth}
\centering
\includegraphics[width=\textwidth]{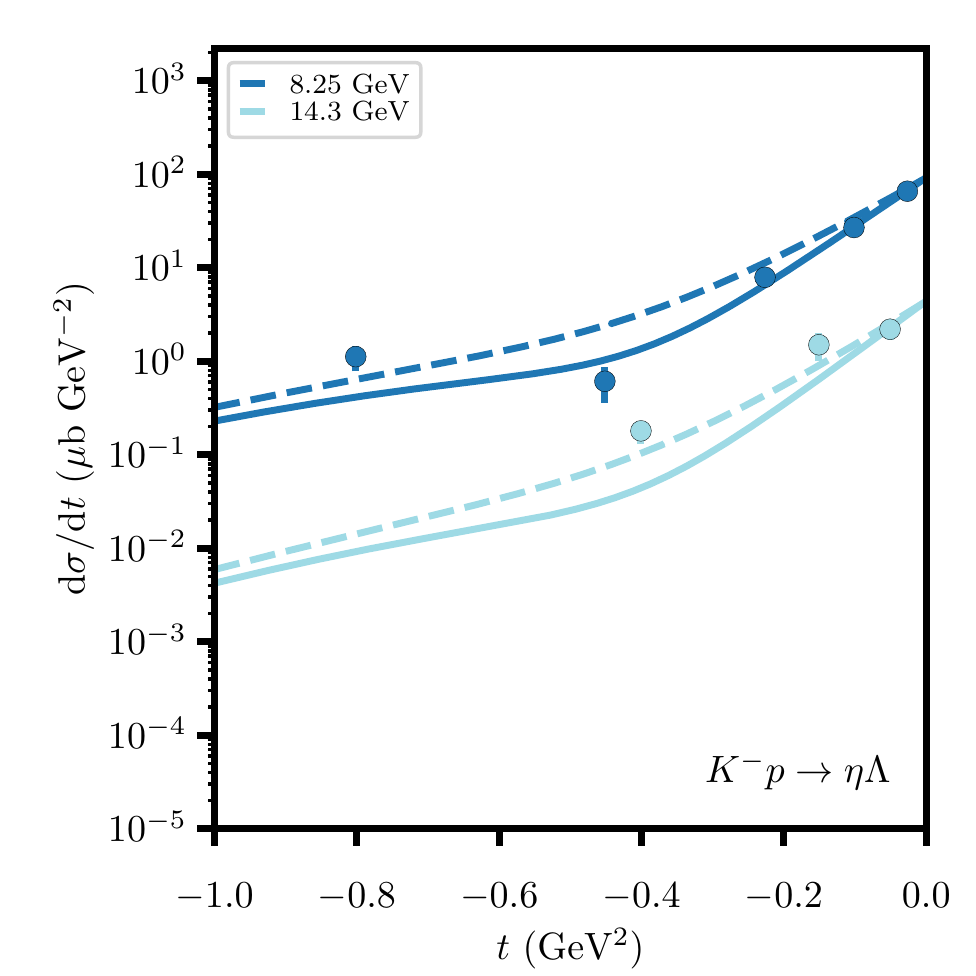}
  \caption{$K^- p \to \eta \Lambda$ \label{fig:K-_P_ETA_LAMBDA}}
\end{subfigure}%
\begin{subfigure}{.5\textwidth}
\centering
\includegraphics[width=\textwidth]{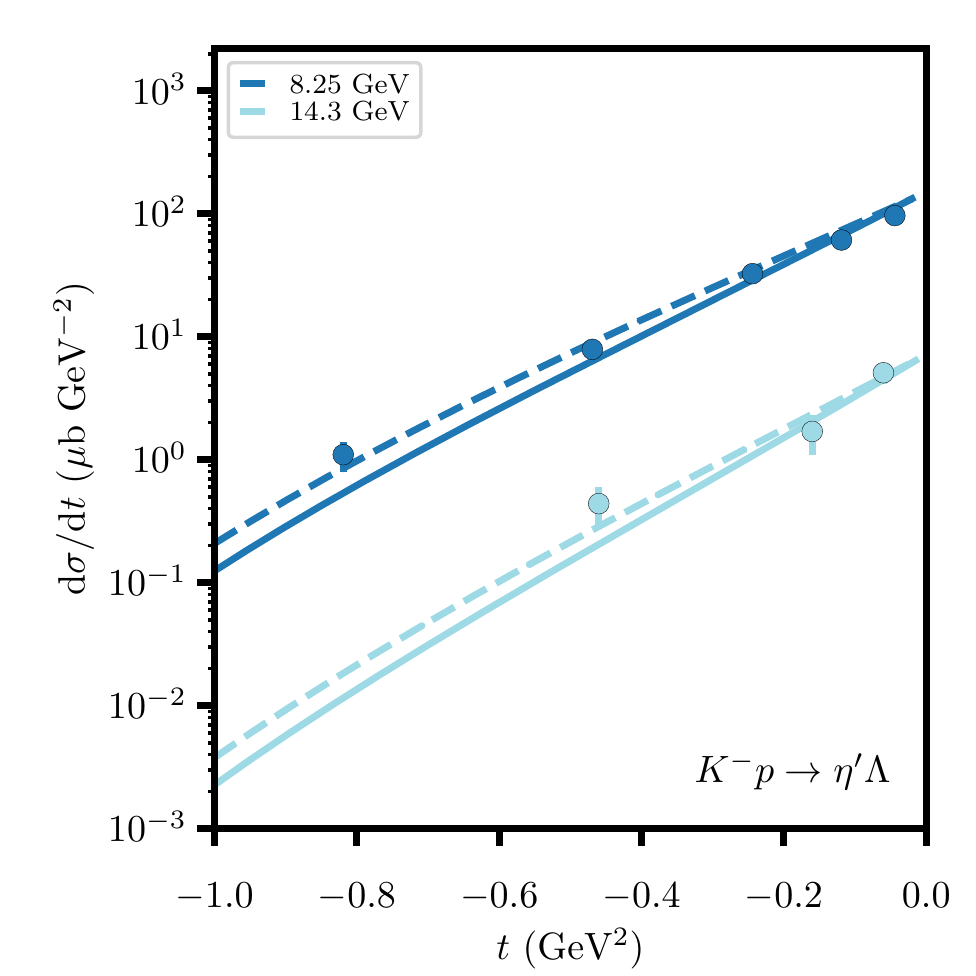}
  \caption{$K^- p \to \eta' \Lambda$ \label{fig:K-_P_ETAPRIME_LAMBDA}}
\end{subfigure}%
\caption{Results for the strangeness exchange $\eta^{(\prime)} \Lambda$ production channels. Scaling and conventions are  as in Fig.~\ref{fig:PI_PI_RHO}.\label{fig:ETAPLAMBDA}}
\end{figure*}

The $\betahat^{K^*_{(2)} K \eta^{(\prime)}}$ couplings are determined in a separate fit using the $\eta^{(\prime)}\Lambda$ production data in Figs.~\ref{fig:K-_P_ETA_LAMBDA}-\ref{fig:K-_P_ETAPRIME_LAMBDA}. As mentioned before, the SU(3) constraint does not hold well for these couplings. 

For the $\Sigma^{*}$ production channels in Figs.~\ref{fig:K-_P_PI-_SIGMA_1385P13_+}-\ref{fig:PI+_P_K+_SIGMA_1385P13_+}, we fix the $\betahat^{K^*_{(2)} N \Sigma^*}_{+\frac{1}{2}+\frac{1}{2}}$ coupling constants and the exponential $t$-dependence to the ones obtained in the global SU(3)-EXD fit. EXD is not imposed for the remaining coupling constants. The SU(3)-EXD fit already provided a reliable representation of these channels. The main difference to the unconstrained fit is the forward behavior of the cross section. Indeed, as discussed before, the single-meson exchange approximation forces the non-flip component to vanish at $t=0$. Therefore, no amplitude survives in the forward direction. In the unconstrained fit, the non-flip component is allowed to contribute in the forward direction. This feature seems to be favored by the data.

\begin{table}
\centering
\caption{Couplings $\betahat^{eif}_{\mu_i \mu_f}(t = m_e^2)$ from the SU(3)-EXD fits. Subscripts denote helicities. The fixed residues are indicated with an asterisk. The exponential factors are not included to extrapolate to the pole. A global fit yields $b_\text{nf} = 1.31\gev^{-2}$ and $b_\text{sf} = 0.54\gev^{-2}$. \label{tab:SU3_residues_pole}}
\begin{tabular}{|C||C|C|C|C|}
\hline
i f				& \rho		& a_2		& K^{*}		& K^*_2	 \\
\hline\hline
\pi^- \pi^- 	& 8.40^*		& .			& .			& .			\\
K^+ K^+ 		& -4.20^*		& 4.20^*		& .			& .			\\
\pi^0 \eta		& .			& 5.78		& .			& .			\\
\pi^0 \eta'		& .			& 6.10		& .			& .			\\
K^- \pi^-		& .			& .			& -5.94^*		& 5.94^*		\\
\kbarz \eta		& .			& .			& -7.20		& -3.48		\\
\kbarz \eta'	& .			& .			& 1.04		& 7.13		\\
\pi^0 \omega_0 	& 0^*			& .			& .			& .			\\
\pi^0 \omega_+	& -15.88		& .			& .			& .			\\
\hline
p_{+} p_{+}				& 1.06		& 1.06		& .			& .			\\	
p_{-} p_{+}				& 5.63		& 5.63		& .			& .			\\	
n_{+} \Lambda_{+}		& .			& .			& -3.33		& -3.33		\\	
n_{-} \Lambda_{+}		& .			& .			& -5.02		& -5.02		\\	
p_{+} \Sigma_{+}^+		& .			& .			& -5.16		& -5.16		\\	
p_{-} \Sigma_{+}^+		& .			& .			& 2.19		& 2.19		\\	
\hline
p_- \Delta^+_{+\frac{1}{2}} 	& 4.08		& 4.08		& .			& .			\\
p_+ \Delta^+_{+\frac{3}{2}} 	& 9.27		& 9.27		& .			& .			\\	
p_+ \Delta^+_{+\frac{1}{2}} 	& 1.70		& 1.70		& .			& .			\\
p_- \Delta^+_{+\frac{3}{2}} 	& 0^*			& 0^*			& .			& .			\\	
p_- \Sigma^{*+}_{+\frac{1}{2}} 	& .			& .			& -2.39		& -2.39		\\
p_+ \Sigma^{*+}_{+\frac{3}{2}} 	& .			& .			& -6.12		& -6.12		\\	
p_+ \Sigma^{*+}_{+\frac{1}{2}} 	& .			& .			& -1.88		& -1.88		\\
p_- \Sigma^{*+}_{+\frac{3}{2}} 	& .			& .			& 0^*	    & 0^*		\\
\hline
\end{tabular}
\end{table}

\begin{table}
\centering
\caption{Couplings $g^{eif}_{\mu_i \mu_f}$ from the unconstrained fits. Subscripts denote helicities. Residues that have been kept fixed in the fit are denoted by an asterisk. \label{tab:unconstrained_residues}}
\begin{tabular}{|C||C|C|C|C|}
\hline
i f				& \rho		& a_2		& K^{*}		& K^*_2	 \\
\hline\hline
\pi^- \pi^- 	& 8.40^*		& .			& .			& .			\\
K^+ K^+ 		& -3.93		& 3.93		& .			& .			\\
\pi^0 \eta		& .			& 5.43		& .			& .			\\
\pi^0 \eta'		& .			& 3.93		& .			& .			\\
K^- \pi^-		& .			& .			& -5.94^*		& 5.94^*		\\
\kbarz \eta		& .			& .			& -7.17		& -2.35		\\
\kbarz \eta'	& .			& .			& 0			& 6.34		\\
\pi^0 \omega_0 	& 0^*			& .			& .			& .			\\
\pi^0 \omega_+	& -9.46		& .			& .			& .			\\
\hline
p_{+} p_{+}				& 1.76		& 1.43		& .			& .			\\	
p_{-} p_{+}				& 8.02		& 7.59		& .			& .			\\	
n_{+} \Lambda_{+}		& .			& .			& -3.77		& -3.77		\\	
n_{-} \Lambda_{+}		& .			& .			& -4.31		& -4.31		\\	
p_{+} \Sigma_{+}^+		& .			& .			& -5.05		& -5.05		\\	
p_{-} \Sigma_{+}^+		& .			& .			& 2.79		& 2.79		\\	
\hline
p_- \Delta^+_{+\frac{1}{2}} 	& 6.26		& 3.22		& .			& .			\\
p_+ \Delta^+_{+\frac{3}{2}} 	& 14.23		& 7.32		& .			& .			\\	
p_+ \Delta^+_{+\frac{1}{2}} 	& 1.25		& -1.83		& .			& .			\\
p_- \Delta^+_{+\frac{3}{2}} 	& 0^*			& 0^*			& .			& .			\\	
p_- \Sigma^{*+}_{+\frac{1}{2}} 	& .			& .			& -2.39^*		& -2.39^*		\\
p_+ \Sigma^{*+}_{+\frac{3}{2}} 	& .			& .			& -10.00	& -6.12		\\	
p_+ \Sigma^{*+}_{+\frac{1}{2}} 	& .			& .			& 0			& -0.94		\\
p_- \Sigma^{*+}_{+\frac{3}{2}} 	& .			& .			& 0^*			& 0^*			\\
\hline
\end{tabular}
\end{table}

\begin{table}
\centering
\caption{Residue exponential factors $b^{eif}_{\mu_i \mu_f}$ from the unconstrained fits. Subscripts denote helicities. The fits are only sensitive to the product of the top and bottom residues.  \label{tab:unconstrained_exponentials}}
\begin{tabular}{|C|C|C|}
\hline
e13			 & e24								& b^{e13}_{\mu_1 \mu_3} + b^{e24}_{\mu_2 \mu_4} (\gev^{-2}) \\
\hline \hline
\rho \pi \pi & \rho N_{+} N_{+}	& 0	\\
\rho \pi \pi & \rho N_{-} N_{+}	& 0.86	\\
a_2 \pi \eta^{(\prime)} & a_2 N_{+} N_{+}	& 0	\\
a_2 \pi \eta^{(\prime)} & a_2 N_{-} N_{+}	& 0.27	\\
\rho K K & \rho N_+ N_+	& -0.55	\\
\rho K K & \rho N_- N_+	& 0.32	\\
a_2 K K & a_2 N_+ N_+	& 0.01	\\
a_2 K K & a_2 N_- N_+	& 0.28	\\
\rho \pi \omega_{+1} & \rho N_+ N_+	& 0.82	\\
\rho \pi \omega_{+1} & \rho N_- N_+	& 1.68	\\
\hline
\rho \pi \pi & \rho N_+ \Delta_{-\frac{1}{2}}	& 1.38	\\
\rho \pi \pi & \rho N_+ \Delta_{+\frac{3}{2}}	& 1.38	\\
\rho \pi \pi & \rho N_+ \Delta_{+\frac{1}{2}}	& 1.85	\\
a_2 \pi \eta^{(\prime)} & a_2 N_+ \Delta_{-\frac{1}{2}}	& -0.16	\\
a_2 \pi \eta^{(\prime)} & a_2 N_+ \Delta_{+\frac{3}{2}}	& -0.16	\\
a_2 \pi \eta^{(\prime)} & a_2 N_+ \Delta_{+\frac{1}{2}}	& 1.30	\\
\rho K K & \rho N_+ \Delta_{-\frac{1}{2}}	& 0.83	\\
\rho K K & \rho N_+ \Delta_{+\frac{3}{2}}	& 0.83	\\
\rho K K & \rho N_+ \Delta_{+\frac{1}{2}}	& 1.31	\\
a_2 K K & a_2 N_+ \Delta_{-\frac{1}{2}}	& -0.15	\\
a_2 K K & a_2 N_+ \Delta_{+\frac{3}{2}} & -0.15	\\
a_2 K K & a_2 N_+ \Delta_{+\frac{1}{2}}	& 1.31	\\
\hline 
K^*_{(2)} K \pi & K^*_{(2)} N_+ \Sigma_+	& 1.26 \\
K^*_{(2)} K \pi & K^*_{(2)} N_+ \Sigma_-	& 0.54 \\
K^*_{(2)} K \pi & K^*_{(2)} N_+ \Lambda_+	& 1.31 \\
K^*_{(2)} K \pi & K^*_{(2)} N_+ \Lambda_-	& 0.54 \\
\hline
K^{*}_{(2)} K \eta^{(\prime)} & K^{*}_{(2)} N_{+} \Lambda_{+}	& 0	\\
K^{*}_{(2)} K \eta^{(\prime)} & K^{*}_{(2)} N_{+} \Lambda_{+}	& 0	\\
\hline
K^{*}_{(2)} K \pi & K^{*}_{(2)} N_+ \Sigma^*_{-\frac{1}{2}}	& 0.54	\\
K^{*}_{(2)} K \pi & K^{*}_{(2)} N_+ \Sigma^*_{+\frac{3}{2}}	& 0.54	\\
K^{*}_{(2)} K \pi & K^{*}_{(2)} N_+ \Sigma^*_{+\frac{1}{2}}	& 1.31	\\
\hline
\end{tabular}
\end{table}

\section{Conclusions}\label{sec:conclusions}
We assessed the applicability of the Regge pole model by performing a global fit to charge and strange exchange quasi-two-body reactions at large momenta $\plab \geq 5\gev$.
We have found that the Regge pole model provides a good description of the data for a large amount of channels, while requiring only a small number of free SU(3) and EXD related parameters. It was shown that the inclusion of these constraints offers a solid way to reduce the number of free parameters of the fit. The large number of free parameters in the unconstrained fit allows for too much freedom compared with the number of available data to yield a unique result. SU(3) and EXD constraints are especially useful to determine the relevant regions of the vast parameter space of the residues. In kinematic domains where wrong-signature zeros can be expected (such as in channels dominated by  $\rho$ exchange), secondary contributions become relevant. For these channels, we find the single pole model and factorization to work well in the domain $0 \leq -t \leq 0.5\gev^2$. The cross sections of reactions dominated by $a_2$ exchanges are well reproduced by the Regge pole models up to $t \sim -0.8\gev^2$, due to the lack of a dip. Reactions dominated by strangeness exchanges follow the Regge pole model remarkably well in the forward region up to $t \sim -0.6\gev^2$. Especially those channels dominated by non-flip baryon vertices are in good agreement with the constraints imposed by SU(3), which should therefore be considered in future fits. These are the kinematic domains were factorization can be used as a reliable approximation to model beam-target fragmentation. The presented model provides a solid description of the production mechanism needed to describe the production amplitude in peripheral resonance production, of relevance to hybrid searches. Our predictions and our model will be made available online on the JPAC website~\cite{JPACweb, Mathieu:2016mcy}. With the online version of the model, users have the possibility to vary the model parameters and generate the observables.

\begin{acknowledgments}
We dedicate this work to the memory of Mike Pennington. 
He was an example of humanity and scientific dedication whose efforts and support made the JPAC Collaboration possible.
This work was supported by 
Research Foundation -- Flanders (FWO),
the U.S.~Department of Energy under grants 
No.~DE-AC05-06OR23177 and No.~DE-FG02-87ER40365,
U.S.~National Science Foundation 
under award numbers PHY-1415459, PHY-1205019, and  PHY-1513524,
PAPIIT-DGAPA (UNAM, Mexico) grant No.~IA101717,
CONACYT (Mexico) grant No.~251817.
\end{acknowledgments}

\FloatBarrier
\clearpage

\makeatletter\onecolumngrid@push\makeatother

\FloatBarrier
\clearpage
\makeatletter\onecolumngrid@pop\makeatother

\appendix

\section{Kinematics and conventions\label{sec:conventions}}
In the $s$-channel center-of-mass (c.m.) frame, the cosine of the scattering angle for $1 + 2 \to 3 + 4$ is given by
\begin{align}\label{eq:z_s}
z_s =& \frac{s^2 + s(2t - \sum_i m_i^2) + (m_1^2 - m_2^2)(m_3^2 - m_4^2)}{S_{12}(s)S_{34}(s)} \,,\\ 
S^2_{ij}(s) \equiv & \left[s - (m_i + m_j)^2\right]\left[s - (m_i - m_j)^2\right]\,.
\end{align}
The conventions are illustrated in Fig.~\ref{fig:kinematics_schematic}.
Introducing $t_{\textrm{min}}(s) \equiv t(z_s = +1)$ and $t' = t - t_{\textrm{min}}$, the above can be simplified
\begin{align}\label{eq:z_s_tprime}
z_s = 1 + \frac{2st'}{S_{12}(s)S_{34}(s)}\,,
\end{align}
or in other words $1-z_s \sim -t'$. 
At leading $s$, $t = t'$.
Furthermore,
\begin{align}
E_i &= \frac{s + m_i^2 - m_j^2}{2 \sqrt{s}} \,, \\
\abs{\vec{p}_i} &= \frac{S_{ij}(s)}{2 \sqrt{s}}\,,
\end{align}
where $(i,j)$ is a $s$-channel pair. 

The expressions for the helicity amplitudes in terms of the covariant couplings are obtained by properly selecting the polarization angles. We use the `particle 2' convention of Jacob and Wick~\cite{Jacob:1959at}, which requires that for the helicity states of the two particles in a two-particle $s$-channel helicity state
\begin{align}
\lim_{\vec{p} \to 0} \braket{-\vec{p},-\mu|\vec{p},\mu} = 1.
\end{align}
Since the baryons in our convention are a `particle 2' in the $s$-channel helicity pairs, one must use the following conventions
\begin{align}
u^{(2)}(p,\mu) = \sqrt{E + m} 
\begin{pmatrix}
\unitm \\
\frac{\vec{p}\cdot \vec{\sigma}}{E+m}
\end{pmatrix}
\chi^{(2)}_\mu(\theta, \phi),
\end{align}
where $\chi^{(2)}(\theta, \phi)$ is a `particle 2' Pauli spinor, i.e.\ the fermion (in a two-particle pair) goes opposite to the direction determined by $(\theta, \phi)$.

For a `particle 1' Pauli spinor we have
\begin{subequations}
\begin{align}
\chi^{(1)}_{+ \frac{1}{2}}(\theta,\phi) &= 
\begin{pmatrix}
\cos \theta \\
e^{i\phi} \sin \theta
\end{pmatrix} \,,\\
\chi^{(1)}_{- \frac{1}{2}}(\theta,\phi) &= 
\begin{pmatrix}
-e^{-i\phi} \sin \theta \\
\cos \theta 
\end{pmatrix}.
\end{align}
\end{subequations}
Using the above definition, the `particle 2' Pauli spinor is defined as
\begin{align}
\chi^{(2)}_{-\mu}(\pi - \theta,\pi+\phi) = \chi^{(1)}_\mu(\theta,\phi).
\end{align}
Also, we consider $\phi = 0$ from hereon. The rotation $\vec{p} \to -\vec{p}$ then corresponds to $(\theta,\phi=0)\to (\pi+\theta,\phi=0)$.
For the produced massive vector meson, we use the polarization vectors
\begin{subequations}
\begin{align}
\epsilon^\nu(p,\mu=\pm 1)&=\frac{1}{\sqrt{2}} \left( 0, -\mu\cos\theta,-i,\mu\sin\theta \right)^T \,,\\
\epsilon^\nu(p,\mu=0)&= \frac{1}{m}\left(\abs{\vec{p}},p^0 \vec{e}_{\vec{p}}\right)^T,
\end{align}
\end{subequations}
where $\vec{e}_{\vec{p}}$ is the unit vector in the direction of $\vec{p}$. For a `particle 2' vector, one finds
\begin{align}
\epsilon^{(2)\nu}(p,\mu)= - g^{\nu \nu } \epsilon^{\nu}(\tilde{p},-\mu)\,,
\end{align}
where the index $\nu$ is not summed over.
For the spin-3/2 Rarita-Schwinger spinor, we use the expression
\begin{align}
u^{(1)\nu}(p,\mu) = \sum_{\mu_1,\mu_2} \braket{1,\mu_1;\frac{1}{2} \mu_2 | \frac{3}{2} \mu} \epsilon^{\nu} (p,\mu_1) u^{(1)}(p,\mu_2) .
\end{align}
since $\vec{p}$ lies in the $x-z$ scattering plane.  Here, $\tilde{p} = (p^0, -\vec{p})$ and $-\vec{p}$ has spherical angles $(\pi-\theta,\phi+\pi)$. 
For a `particle 2' spin-3/2 spinor, we use the same form, with the spinor and polarization vector substituted by their `particle 2' form. In the following, we drop the explicit `particle 1' and `particle 2' reference for brevity of notation. 
The spin-3/2 spinors satisfy the Rarita-Schwinger equations
\begin{align}
p_\nu u^{\nu}(p,\mu) &= 0 \,,\\
\gamma_\nu u^{\nu}(p,\mu) &= 0 \,,\\
(\slashed{p} - m) u^{\nu}(p,\mu) &= 0. 
\end{align}

Within our conventions, parity invariance implies
\begin{align}
A_{\mu_4 \mu_3 \mu_2 \mu_1} &= \eta^s A_{-\mu_4 -\mu_3 -\mu_2 -\mu_1}\,, \\
\eta^s &= \frac{\eta_1 \eta_2}{ \eta_3 \eta_4}  (-1)^{\mu' - \mu},
\end{align}
where $\eta_i = P_i (-1)^{s_i}$ is the naturality of a particle $i$ with parity $P_i$ and spin $s_i$.
For the individual vertices, one has
\begin{align}\label{eq:parity_residues}
\beta_{\mu_3 \mu_1} &= \eta_e P_1 P_3 (-1)^{s_3 - s_1} (-1)^{\mu_3 - \mu_1} \beta_{-\mu_3 -\mu_1} \,,\\
\beta_{\mu_4 \mu_2} &= \eta_e P_2 P_4 (-1)^{s_4 - s_2} (-1)^{\mu_2 - \mu_4} \beta_{-\mu_4 -\mu_2} \,,
\end{align}
where $\eta_e$ is the naturality of the exchange.

\section{Factorization of Regge pole residues}\label{sec:factorization}
Since testing factorization is the central topic of this work, we derive its implications on helicity amplitudes. The kinematic $t$-singularities in the $s$-channel partial waves are given by the half-angle factors in Eq.~\eqref{eq:half_angle_factor}.
Hence, for $z_s \to +1$ (or equivalently $t' \to 0$), one finds
\begin{align}\label{eq:half_angle_tprime}
A_{\mu_4 \mu_3 \mu_2 \mu_1} \sim \sqrt{-t'}^{\abs{\mu - \mu'}}  = \sqrt{-t'}^{\abs{\omega - \omega'}},
\end{align}
where $\omega = \mu_1 - \mu_3$ and $\omega' = \mu_2 - \mu_4$. 
The above is the most singular kinematic behavior of the amplitude, and cannot be cast into the factorizable form in Eq.~\eqref{eq:factorization}. One can show~\cite{Frampton:1968rw} that the simplest factorizable form consistent with a definite parity exchange is given by 
\begin{align}
A_{\mu_4 \mu_3 \mu_2 \mu_1} \sim \sqrt{-t}^{\abs{\omega}+\abs{\omega'}}.
\end{align}
In Eq.~\eqref{eq:general_regge_form}, $t'$ is used explicitly for the factors stemming from the half-angle factors in Eq.~\eqref{eq:half_angle_tprime}.

\section{Interaction Lagrangians\label{sec:lagrangians}}

\begin{table}[tb]
\def\arraystretch{1.8}
\caption{Listing of the reduced $s$-channel residues and their expressions in the single-particle exchange formalism. The expressions are given up to the flavor trace. We assumed $M_{S} = (m_i +m_f)/2$. In practice, these couplings are evaluated on the mass pole of the exchanged particle. Only the lowest order $t$-dependence, which is compatible with factorization, is used in the residue. Every residue contains an additional $\sqrt{s_0}^{J_e}$, where $J_e$ is the spin of the lightest particle on the trajectory. \label{tab:s_channel_residues}}
\begin{tabular}{|C|C|}
\hline
\betahat^{R i f}_{\mu_i \mu_f}(t) & \text{Expression}    \\
\hline \hline
\betahat^{VPP}_{00} & \sqrt{2}\, g_{VPP}  \\
\hline
\betahat^{VBB}_{+\frac{1}{2}+\frac{1}{2}} & \sqrt{2}\, g^v_{VBB}  \\
\betahat^{VBB}_{-\frac{1}{2}+\frac{1}{2}} & \sqrt{2}\, g^t_{VBB} / \left( m_2 + m_4\right)   \\
\hline
\betahat^{TPP}_{00} & g_{TPP} / 2  \\
\hline
\betahat^{TBB}_{+\frac{1}{2}+\frac{1}{2}} & - \frac{2}{m_2 + m_4} \left(g_{TBB}^{(2)}  +  g_{TBB}^{(1)} \right)  \\
\betahat^{TBB}_{-\frac{1}{2}+\frac{1}{2}} &  2 \, g_{TBB}^{(2)}/(m_2 + m_4)^2   \\
\hline
\betahat^{VBD}_{-\frac{1}{2} +\frac{1}{2}} & 2\frac{g^{(1)}_{VBD} m_2(m_2 + m_4) - g^{(2)}_{VBD} (2t - m_4 (m_2 - m_4)) + 2t g^{(3)}_{VBD}}{\sqrt{3} m_4 (m_2 + m_4)^2}   \\
\betahat^{VBD}_{+\frac{1}{2} +\frac{3}{2}} & \frac{2}{(m_2 + m_4)^2}\left[g^{(1)}_{VBD} (m_2+m_4) - g^{(2)}_{VBD} (m_2 - m_4)\right]\\
\betahat^{VBD}_{+\frac{1}{2}+\frac{1}{2}} & \frac{g^{(1)}_{VBD}(m_2 + m_4) - g^{(2)}_{VBD} (m_2 - m_4) - g^{(3)}_{VBD} (m_2 - m_4)}{\sqrt{12} m_4 (m_2 + m_4)} (-t)   \\
\betahat^{VBD}_{-\frac{1}{2} +\frac{3}{2}} & 2 \, g^{(2)}_{VBD} / (m_2 + m_4)^2  \\
\hline
\betahat^{VPV}_{00} & 0~(\text{Parity}) \\
\betahat^{VPV}_{0+1} & g_{VVP}/2   \\
\hline 
\betahat^{AVP}_{0,0}(t) &  g_{AVP} / \left( \sqrt{2}\, m_3 \right)  \\
\betahat^{AVP}_{0+1}(t) &  g_{AVP}'   \\
\hline
\betahat^{ABB}_{+\frac{1}{2}+\frac{1}{2}} & 
0~(\text{for }C_A=-1)  \\
\betahat^{ABB}_{-\frac{1}{2}+\frac{1}{2}} & \sqrt{2}\, g^t_{ABB}/(m_2 + m_4) \\
\hline
\betahat^{TPV}_{00} & 0 ~(\text{Parity})   \\
\betahat^{TPV}_{0+1} & g_{TPV} / \sqrt{2}   \\
\hline
\end{tabular}
\end{table}

In order to relate the helicity couplings to those of modern literature, we start from the set of effective Lagrangians given below. We consider interactions of pseudoscalars (P), vectors (V), axial vectors (A), tensors (T), 
octet baryons (B) and decuplet baryons (D).

We use the following conventions in the global SU(3) fit. The explicit form of the matrices for pseudoscalar, vector and tensor mesons is
\begin{align}
\nn P&=\left(\begin{array}{ccc}
\frac{\pi^0}{\sqrt{2}}+\frac{\eta^8}{\sqrt{6}}+\frac{\eta^1}{\sqrt{3}} & \pi^+ & K^+\\
\pi^- & \frac{-\pi^0}{\sqrt{2}}+\frac{\eta^8}{\sqrt{6}}+\frac{\eta^1}{\sqrt{3}} & K^0\\
K^- & \bar{K}^0 & -\sqrt{\frac{2}{3}}\eta^8+\frac{\eta^1}{\sqrt{3}}
\end{array}\right),\\
\nn \eta^1&=\eta'\cos\theta_P-\eta\sin\theta_P,\\
\eta^8&=\eta\cos\theta_P+\eta'\sin\theta_P,\label{eq:P_matrix}\\[1em]
\nn V^8 &=
\left( 
\begin{array}{ccc}
\frac{1}{\sqrt{2}} \rho^0 + \frac{1}{\sqrt{6}} \omega^8 & \rho^+ & K^{\ast +} \\[1ex] 
\rho^- & -\frac{1}{\sqrt{2}} \rho^0 + \frac{1}{\sqrt{6}} \omega^8 & K^{\ast 0} \\[1ex] 
K^{\ast -} & \bar K^{\ast 0} & -\frac{2}{\sqrt{6}} \omega^8
\end{array}
\right),\\[1em]
 V^1  &=  \omega^1,\\[1em]
\nn T^8&=\left(\begin{array}{ccc}
\frac{a_2}{\sqrt{2}}+\frac{f_2^8}{\sqrt{6}} & a_2^+ & K_2^{*+}\\
a_2^- & -\frac{a_2^0}{\sqrt{2}}+\frac{f_2^8}{\sqrt{6}} & K_2^{*0}\\
K_2^{*-} & \bar{K}_2^{*0} & -2\frac{f_2^8}{\sqrt{6}}
\end{array}\right),\\[1em]
\nn T^1&=f_2^1=f_2\cos\theta_T-f_2'\sin\theta_T,\\
f_2^8&=f_2\sin\theta_T+f_2'\cos\theta_T.
\end{align}
Here, $\theta_T$ and $\theta_P$ are the tensor and pseudoscalar mixing angles, respectively, between singlet and octet. Furthermore, we call $\theta_P^I$ the ideal mixing angle with $\sin\theta_P^I=1/\sqrt{3}$ and $\cos\theta_P^I=\sqrt{2/3}$. In the case of ideal mixing for vector mesons, the following relations hold:
\begin{align}
\omega= \frac{1}{\sqrt{3}} \omega^8 + \sqrt{\frac{2}{3}} \omega^1,\quad\phi= \sqrt{\frac{2}{3}} \omega^8 - \frac{1}{\sqrt{3}} \omega^1.
\end{align}\\[-.5em]

Note that we have implicitly assumed a nonet symmetry for the pseudoscalar mesons in Eq.~\eqref{eq:P_matrix}, where the singlet/octet coupling ratio is $S_T = \sqrt{2}$~\cite{Fox:1973by}. The latter corresponds to neglecting the coupling to $s\bar s$ content. In the following, the trace is taken over flavor space, corresponding to the isospin couplings of the different channels in SU(3) symmetry.

The three-meson interaction Lagrangians consist only of a symmetric coupling due to $G$-parity conservation. All the couplings appearing in the following are fitting parameters, obtaining values as explained in App.~\ref{sec:couplings_estimations}. The Lagrangians describing the couplings of tensor mesons to pseudoscalar and vector mesons are given by~\cite{Bellucci:1994eb,Chow:1997sg,Giacosa:2005bw}
\begin{align}
\mathcal{L}_{TPP}&=c_{TPP}^8\langle\mathcal{T}^8_{\mu\nu}\Theta_P^{\mu\nu}\rangle+\frac{c_{TPP}^1}{\sqrt{3}}\mathcal{T}^1_{\mu\nu}\langle\Theta_P^{\mu\nu}\rangle,\label{EqLagTPP}\\
\mathcal{L}_{TVP} &=g_{TVP}\left<\mathcal{T}^{[\mu\nu]\alpha}\left[\tilde{V}_{\mu\nu},\partial_\alpha P\right]\right>,\label{eq:L_TVP}
\end{align}
where
\begin{align}
\Theta_P^{\mu\nu}&=\partial^\mu P\partial^\nu P - g^{\mu\nu}(\partial\cdot P)^2,\label{ETheta}\\
\mathcal{T}^{[\mu\nu]\alpha}&=\partial^{\mu}\mathcal{T}^{\nu\alpha}-\partial^{\nu}\mathcal{T}^{\mu\alpha},\\
\mathcal{T}_{\mu\nu}&=\mathcal{T}^8_{\mu\nu}+\frac{\mathcal{T}^1_{\mu\nu}}{\sqrt{3}},\\
\tilde{V}_{\mu\nu}&=\frac{1}{2}\epsilon_{\mu\nu\rho\sigma}V^{\rho\sigma},\\
V^{\mu\nu}&=\partial^\mu V^\nu-\partial^\nu V^\mu.
\end{align}
Concerning the couplings of vector to pseudoscalar mesons, the Lagrangians read
\begin{align}
\mathcal{L}_{VPP} &= -\mathrm{i}g_{VPP}\langle \left[P,\partial_\mu P\right] V^\mu\rangle, \label{eq:L_VPP}\\[.5em]
\mathcal{L}_{VVP} &=\frac{g_{VVP}}{2}\epsilon_{\mu\nu\alpha\beta}\left<\left\{\partial^\mu V^\nu,\partial^\alpha V^\beta\right\} P\right>.\label{eq:L_VVP}
\end{align}

The Lagrangian that couples vector, pseudoscalar and axial mesons $b_1$ is given by (we do not consider SU(3) relations for the $b_1$ exchange, since it is the only axial exchange considered here)~\cite{Roca:2005nm,Roca:2006am}.
\begin{equation}
\mathcal{L}_{VPA}=g_{VPA}A_\mu V^\mu P + g_{VPA}' A_{\mu \nu} V^{\mu \nu} P \,.
\end{equation}

Since one can couple two octets to both a symmetric and an antisymmetric octet, each meson--$B$--$B$ vertex must be decomposed into two irreducible structures, with independent couplings. The couplings of the vector-meson fields $V_\mu$ with momentum $q$ to the octet baryons are described by the following Lagrangian~\cite{Drechsel:1998hk}
\begin{widetext}
\begin{align}
\nonumber\mathcal{L}_{VBB}=\Bigg\langle \bar{B}\Bigg[&\Big(g^{v,F}_{VBB} \left[V_\mu^8, B\right]+g^{v,D}_{VBB} \left\{V_\mu^8, B\right\}
+g^{v,S}_{VBB} V_\mu^1 B\Big)\gamma^\mu\\
+&\Big(g^{t,F}_{VBB} \left[V_\mu^8, B\right]+g^{t,D}_{VBB} \left\{V_\mu^8, B\right\}+g^{t,S}_{VBB} V_\mu^1 B\Big)\frac{\mathrm{i}\sigma^{\mu\nu}q_\nu}{2M_S}\Bigg]\Bigg\rangle ,\label{eq:L_VNN}
\end{align}
\end{widetext}
where $M_S=(m_{B1}+m_{B2})/2$, and $m_{B1}$ and $m_{B2}$ are the masses of the incoming and the outgoing baryon, respectively. The octet-baryon matrix is explicitly given as
\begin{align}
B&=\left(\begin{array}{ccc}
\frac 1{\sqrt 2}\Sigma^0+\frac1{\sqrt 6}\Lambda& \Sigma^+ &p\\
\Sigma^-& -\frac1{\sqrt 2}\Sigma^0+\frac1{\sqrt6}\Lambda& n\\
\Xi^-&\Xi^0&-\frac 2 {\sqrt6}\Lambda
\end{array}\right).
\end{align}\\[-.5em]

In the following, for simplicity we drop the notation with traces over commutators/anticommutators in flavor space. The couplings of tensor mesons $f^{\mu\nu}$ to the baryons are given in Refs.~\cite{Kleinert:1971qg,Yu:2011zu,Oh:2003aw,Yu:2011fv,Huang:2016gdt}.
We follow the general definitions of the Lagrangian
\begin{widetext}
\begin{align}
\mathcal{L}_{TBB}&=\mathrm{i}\frac{g_{TBB}^{(1)}}{4M_S}\bar{B}\left(\gamma_\mu\overleftrightarrow{\partial}_\nu+\gamma_\nu\overleftrightarrow{\partial}_\mu\right)Bf^{\mu\nu}+\frac{g_{TBB}^{(2)}}{M_S^2}\partial_\mu\bar{B}\partial_\nu Bf^{\mu\nu}, \label{eq:L_TBB}
\end{align}
\end{widetext}
where the SU(3) couplings are given by the usual traces in flavor space, analogously to those in Eq.~\eqref{eq:L_VNN}. 

An axial vector couples to octet baryons via
\begin{align}
 \mathcal{L}_{ABB}&=\bar{B}\left(g^v_{ABB}\gamma^\mu+g^t_{ABB}\frac{\mathrm{i}\sigma^{\mu\nu}q_\nu}{2M_S}\right)\gamma_5 A_\mu B. \label{eq:L_ANN}
 \end{align}
The explicit expressions for the flavor couplings are analogous to those in Eq.~\eqref{eq:L_VNN}. Care must be taken in Eq.~\eqref{eq:L_ANN}, since only one of the couplings is allowed for a definite $G$-parity state of the axial meson $A$. Because of $G$-parity considerations, the vector coupling $g^v_{ABB}$ does not contribute to $b_1$ exchanges, and the tensor coupling $g^t_{ABB}$ is blocked for $a_1$ exchanges.

The Lagrangians that describe the octet-to-decuplet transitions via a vector-meson and a tensor-meson emission are~\cite{Nam:2011np,Yu:2016jfi}
\begin{widetext}
 \begin{align}
 \mathcal{L}_{VBD}&=-\sqrt{3}\bar{D}_{ijk}^\mu\epsilon_{ilm}\left[-\mathrm{i}\frac{g_{VBD}^{(1)}}{M_S}\gamma^\nu V_{\mu\nu,jl}
 -\frac{g_{VBD}^{(2)}}{M_S^2}V_{\mu\nu,jl}\partial^\nu 
 +\frac{g_{VBD}^{(3)}}{M_S^2}\partial^\nu V_{\mu\nu,jl}\right]\gamma_5 B_{km}+\text{h.c.},\label{eq:L_VBD}\\
 \mathcal{L}_{TBD}&=-\sqrt{3}\mathrm{i}\frac{g_{TBD}}{m_T}\bar{D}_{ijk}^\lambda\epsilon_{ilm}
 \left(g_{\lambda\mu}\partial_\nu+g_{\lambda\nu}\partial_\mu\right)\gamma_5 B_{km}T^{\mu\nu}_{jl}+\text{h.c.},\label{EqLagVND, eq:L_TBD}
\end{align}
 \end{widetext}
where the explicit form of the fully symmetric decuplet matrix elements is
\begin{align}
T_\mu^{111}=&\Delta_\mu^{++}\,, \ 
   T_\mu^{112}=\frac1{\sqrt3}\Delta_\mu^{+}\,, \nonumber \\ 
   T_\mu^{122}=&\frac1{\sqrt3}\Delta_\mu^{0}\,, \ 
   T_\mu^{222}=\Delta_\mu^{-}\,,\nonumber\\[1em]
T_\mu^{113}=&\frac1{\sqrt3}\Sigma_\mu^{\ast+}\,, \
   T_\mu^{123}=\frac1{\sqrt6}\Sigma_\mu^{\ast 0}\,, \ 
   T_\mu^{223}=\frac1{\sqrt3}\Sigma_\mu^{\ast -}\,, \nonumber\\[1em]
T_\mu^{133}=&\frac1{\sqrt3}\Xi_\mu^{\ast 0}\,, \ 
   T_\mu^{233}=\frac1{\sqrt3}\Xi_\mu^{\ast -}\,, \
   T_\mu^{333}=\Omega_\mu^{-}.
\end{align}
 Note that in the literature one often does not consider the minimal complete set of interaction Lagrangians when spin-$3/2$ baryons or tensor mesons are involved. In these analyses, one rather selects a single interaction term, which is usually unjustified. When considering $\betahat^{VBD}_{\mu_V \mu_D}$ in Table~\ref{tab:s_channel_residues}, one observes that for $g^{(2)}_{VBD} = g^{(3)}_{VBD} = 0$ and $m_2 = m_4$, and neglecting $t$-dependent terms in $\betahat^{VBD}_{\mu_V \mu_D}(t)$, we reproduce the quark-model results in~\cite{Irving:1977ea}: $\beta^{VBD}_{-\frac{1}{2} +\frac{1}{2}}/\beta^{VBD}_{+\frac{1}{2} +\frac{3}{2}} = \sqrt{3}$ and $\beta^{VBD}_{+\frac{1}{2} +\frac{1}{2}} = \beta^{VBD}_{-\frac{1}{2} +\frac{3}{2}} = 0$. Based on this correspondence, we set  $g^{(2)}_{VBD} = g^{(3)}_{VBD} = 0$ in the global SU(3)-EXD fit.

\section{Estimating coupling constants from decay ratios\label{sec:couplings_estimations}}
The coupling constants that appear in the meson Lagrangians can be estimated based on measured decay ratios. These couplings serve as starting values for the fits.

From the data on decay widths of tensor mesons into pseudoscalars, one can extract the numerical values for the couplings $c_{TPP}^8$ and $c_{TPP}^1$. When doing so in a global fit, Giacosa \etal~\cite{Giacosa:2005bw} obtained\footnote{In~\cite{Giacosa:2005bw}, there was a typo in the results, which was solved in private communication -- the values given in that paper are $c_{TPP}^{i,\text{Paper}}=\frac{\sqrt{2}F^2}{4}c_{TPP}^{i,\text{True}}$, where $F=92.4$~MeV is the pion decay constant.}
\begin{align}
c_{TPP}^8\approx 11.7\gev^{-1},\quad c_{TPP}^1\approx 13.6\gev^{-1}.
\end{align}

The decay width of the $\rho$ into two pions is $\Gamma_{\rho\rightarrow\pi\pi}=(149.1\pm 0.8)~\unit{MeV}$~\cite{pdg}. Comparing this with the expression for the decay width in terms of the Lagrangian couplings, one finds $g_{VPP}=\pm 4.2$. For completeness, it is worth mentioning that $g_{VPP}=\pm 4.5$, when extracted from the $K^*\rightarrow K\pi$ decay width.

Estimating the coupling $g_{VVP}$ is less straightforward. The decay ratio of the $\phi$ meson into the $\omega\pi^0$ channel is consistent with 0, both from theory and from experiment, while the only other channel measured so far is the decay of the $\phi$ into 3 pions, which can occur via the intermediate channel $\phi\rightarrow\rho\pi$. If one were to assume ideal mixing for the vector mesons, this coupling would vanish as well in the OZI limit. Therefore, for this particular estimate, we use the vector-meson mixing angle $\theta_V=39^o$~\cite{Giacosa:2005bw}. Assuming that all 3-pion decays of the $\phi$ happen via the $\pi\rho$ intermediate channel, one then obtains $g_{VVP}=13.2\gev^{-1}$. Note that this is to be seen only as a starting value for the fits, since for this particular value many assumptions had to be made, which can give only a rough estimate of the coupling value.

To estimate $g_{TVP}$, we use the information that the decay $a_2 \to 3\pi$ with decay width $105~\unit{MeV}$ occurs dominantly through the $\rho \pi$ intermediate state. Assuming only this intermediate state, one finds a partial width of $73.5~\unit{MeV}$. This leads to $g_{TVP}\approx 6.8\gev^{-2}$. In fact, the coupling ranges between $6.4\gev^{-2}$ and $6.8\gev^{-2}$, when the decay fraction into a $\rho \pi$ intermediate state is varied between $88\%$ and $100\%$. This is consistent with the results from $K_2^* \to \pi K^*$, $K\rho$ and $K\omega$, where the estimated $g_{TVP}$ ranges between $6.4\gev^{-2}$ and $7.5\gev^{-2}$.

In order to estimate $g_{VPA}$, we use the total decay width $(142\pm 9)~\unit{MeV}$ of the $b_1$. It dominantly decays into $\omega\pi$, leading to $g_{VPA}\approx 4.0\gev$. 
All of the above-mentioned estimates are in good agreement with the quark-model predictions from~\cite{Godfrey:1985xj}.

One can extract the meson-baryon couplings using Eq.~\eqref{eq:L_VNN} and ($i = v,t$)
 \begin{align}
 g^i_{\rho pp}&=\frac{1}{\sqrt{2}}(g^{i,F}_{VBB}+g^{i,D}_{VBB}), \\
 g^i_{\omega pp}&=\frac{3g^{i,F}_{VBB}-g^{i,D}_{VBB}}{3\sqrt{2}}+\sqrt{\frac{2}{3}}g^{i,S}_{VBB},
\end{align}
and relating them to the empirical couplings from nucleon-nucleon scattering data. The empirical results of the nucleon-nucleon Bonn potential from Refs.~\cite{Machleidt:1987hj,Machleidt:2000ge} read
\begin{align}
g^v_{\rho pp}=3.3,\quad g^v_{\omega pp}=16,\quad g^t_{\rho pp}=20,\quad g^t_{\omega pp}=0.
\end{align}
Furthermore, from~\cite{Dover:1985ba}, one finds
\begin{align}
\frac{g_{v}^F}{ g_{v}^F + g_{v}^D } = 1,\quad \frac{g_{v}^F+g_{t}^F}{ g_{v}^F + g_{v}^D + g_{t}^F + g_{t}^D } = \frac{2}{5}.
\end{align}
Finally, one obtains
\begin{align}
\nn&g^{v,F}_{VBB} = 4.6, & g^{v,D}_{VBB} = 0, & \quad g^{v,S}_{VBB} = 15.5, \\[1em]
 &g^{t,F}_{VBB} = 8.4,  & g^{t,D}_{VBB} = 19.6,  & \quad g^{t,S}_{VBB} = -1.6.
\label{Eq:coupVN}
\end{align}

The coupling $g_{TBB}^{(2)}$ in Eq.~\eqref{eq:L_TBB} is often estimated to be compatible with $0$ when tensor-meson dominance (TMD) is assumed. One might question the validity of the TMD approach. Indeed, in~\cite{Kleinert:1971qg} the authors state that $g_{TBB}^{(1)}\approx -g_{TBB}^{(2)}$.

\section{Covariant and helicity amplitudes\label{sec:amplitudes}}
From the interaction Lagrangians in Appendix~\ref{sec:lagrangians}, one determines the helicity amplitudes by choosing the appropriate polarization angles. Since we consider high-energy scattering, the amplitudes are expanded in powers of $s$ and only the leading term is used. The $s$-channel helicity amplitudes are then cast onto the factorized form 
\begin{align}
A_{\mu_4 \mu_3 \mu_2 \mu_1} &= \beta^{e13}_{\mu_1 \mu_3}(t) \beta^{e24}_{\mu_2 \mu_4}(t) \propagator{e}(s,t),
\end{align}
which coincides with the limit of our high-energy amplitudes in Eq.~\eqref{eq:asymptotic_regge_form} for $t\to m_e^2$. We introduced the convenient notation for the meson propagator
\begin{align}
\propagator{e} \equiv \propagator{e}(s,t) &= \frac{s^{J_e}}{m_e^2 - t}.
\end{align}
From the analysis of kinematical singularities, we showed that the $s$-channel residues $\beta^{eij}_{\mu_i \mu_j}(t)$ of an evasive reggeon must at least go as $\beta^{eij}_{\mu_i \mu_j}(t) \sim \sqrt{-t}^{\abs{\mu_i - \mu_j}}$. In order to unambiguously factorize our amplitudes, we first consider $\pi \pi$, $\pi N$ and $N N$ scattering with a $t$-channel vector exchange $V$. One obtains the asymptotic expressions (up to isospin factors)
\begin{subequations}
\begin{align}
A_{0000}(\pi \pi \to \pi \pi) &= 2 g_{VPP}^2 \propagator{V}\,,\\
A_{+0+0}(\pi N \to  \pi N) &= 2 g_{VPP} g^v_{VBB}  \propagator{V}\,,\\
A_{++++}(NN \to NN) &= 2 (g^v_{VBB})^2 \propagator{V} \,,\\
A_{+++-}(NN \to NN) &= g_v \frac{g^t_{VBB}}{M_B} \sqrt{-t'} \propagator{V}\,.
\end{align}
\end{subequations}
From the above, we obtain 
\begin{subequations}
\begin{align}
\beta^{VPP}_{00}(t) &= \sqrt{2} g_{VPP} \\
\beta^{V B B}_{++}(t) &= \sqrt{2}g^v_{VBB}\\
\beta^{V B B}_{-+}(t) &=  \sqrt{2}\frac{g^t_{VBB}}{m_2 + m_4}\sqrt{-t'} .
\end{align}
\end{subequations}
A similar approach is followed for all reactions under consideration in this work. All residues considered in this work are listed in Table~\ref{tab:s_channel_residues}. The SU(3) flavor traces and commutators appear as factors then multiplying these residues.

As an example for how to build an amplitude inspired by single-particle-exchange, one can consider the helicity-flip contribution to the process $\pi^- p \to \pi^0 n$. The full SU(3)-constrained amplitude at high energies is given by
\begin{align}
& A_{+0-0}(\pi^- p \to \pi^0 n) \nonumber \\
&\qquad 
=e^{b_\text{sf}t} \sqrt{-t} \betahat^{\rho \pi^- \pi^0}_{00}(t) \betahat^{\rho p n}_{-+}(t) \mathcal{F}_\rho(s,t) \mathcal{R}(s,t).
\end{align}
Here,
\begin{align}
\betahat^{\rho \pi^- \pi^0}_{00}(t) &= \sqrt{2}\left(\sqrt{2} g_{VPP}\right) \,,  \\
\betahat^{\rho p n}_{-+}(t) &= \frac{\sqrt{2}}{2 m_N}\left(g^{t,F}_{VBB}+g^{t,D}_{VBB}\right),
\end{align}
where the factor in front of the brackets is obtained from Table~\ref{tab:s_channel_residues}. The terms between brackets are obtained by working out the flavor traces. 

In the unconstrained fit, we have
\begin{widetext}
\begin{align}
A_{+0-0}(\pi^- p \to \pi^0 n) &= e^{(b^{\rho \pi \pi}+b^{\rho N N}_{-+})t}\left(- g^{\rho \pi^+ \pi^+}\right)\left( \sqrt{2} g^{\rho p p}_{-+}  \right) \sqrt{-t} \mathcal{F}_\rho(s,t)\mathcal{R}(s,t)\,,
\end{align}
\end{widetext}
since we have made the arbitrary choice of fitting the coupling constants $g^{\rho \pi^+ \pi^+}$ and $\beta^{\rho p p}_{-+}$ and relating all other charge states to these couplings.

The differential cross section is computed using
\begin{align}
\frac{\diffd \sigma}{\diffd t} = \frac{1}{2 \times 16\pi S_{12}^2(s)}\sum\limits_{\mu_i} \abs{A_{\mu_4 \mu_3 \mu_2 \mu_1}}^2.
\end{align}

\section{$t$-channel decay couplings}\label{sec:rotation}
In various single channel analyses, one starts from the $t$-channel to model the reggeon contributions. Hereby, $t$-channel helicity couplings (denoted by $\gamma_{\lambda_i \lambda_k}(t)$) are used. The $s$-channel residues can be related to the $t$-channel residues by analytically continuing the $s$-channel amplitudes to the $t$-channel. To distinguish between the helicity amplitudes in both channels we explicitly mention the channel in a subscript. Additionally, we use $\mu$ and $\lambda$ for $s$- and $t$-channel helicities, respectively.

Consider the $s$-channel amplitude for an exchange $e$ in Eq.~\eqref{eq:asymptotic_regge_form}. Near the lowest mass pole we have the simple form for the $s$-channel amplitude
\begin{align}\label{eq:s_channel_single_pole}
A^s_{\mu_4 \mu_3 \mu_2 \mu_1}(s,t) = \frac{\beta_{\mu_2\mu_4}(t) \beta_{\mu_1 \mu_3}(t)  }{m_e^2-t} s^{J_e}.
\end{align}
Note that in the above, the residues include the $\sqrt{-t}$ factors, which must also be evaluated at $t = m_e^2$.

In the $t$-channel $a + \overline{c} \to \overline{b} + d$, the helicity amplitude can be expanded in partial waves through
\begin{align}\label{eq:pwe_t_ch}
A^t_{\lambda_4 \lambda_2 \lambda_3 \lambda_1}(s,t) = 16 \pi \sum_{J = M_t} (2 J + 1) A^t_{\lambda_4 \lambda_2 \lambda_3 \lambda_1, J}(t) d^J_{\lambda \lambda'}(z_t),
\end{align}
where $\lambda = \lambda_{13} = \lambda_1 - \lambda_3$ and $\lambda' = \lambda_{24} = \lambda_2 - \lambda_4$ and $M_t = \max \{\abs{\lambda},\abs{\lambda'}\}$. The $z_t$ is the cosine of the $t$-channel c.m.\ scattering angle.
For a resonance $e$ with spin $J_e$, the corresponding partial-wave amplitude is parametrized as
\begin{align}\label{eq:pw_pole_tch}
A^t_{\lambda_4 \lambda_2 \lambda_3 \lambda_1, J_e}(t) &=\frac{\gamma_{\lambda_2\lambda_4}(t) \gamma_{\lambda_1 \lambda_3}(t)}{m_e^2-t}.
\end{align}
In the following, we relate the $s$-channel residues $\beta$ in Eq.~\eqref{eq:s_channel_single_pole} to the $t$-channel residues $\gamma$ in Eq.~\eqref{eq:pw_pole_tch}, without invoking Lagrangians to carry out the crossing (which is the inverse direction of the derivation by Fox and Hey~\cite{Fox:1973by}).

We continue the $s$-channel helicity amplitudes into the $t$-channel and project them onto the $t$-channel helicity basis~\cite{Martin:1970}
\begin{widetext}
\begin{align}\label{eq:crossing_matrix}
A^t_{\lambda_4 \lambda_2 \lambda_3 \lambda_1}(s,t) &= -i\sum_{\mu_i} d^{s_1}_{\mu_1 \lambda_1}(-\chi^{t \to s}_1) d^{s_2}_{\mu_2 \lambda_2}(-\chi^{t \to s}_2) d^{s_3}_{\mu_3 \lambda_3}(-\chi^{t \to s}_3) d^{s_4}_{\mu_4 \lambda_4}(-\chi^{t \to s}_4) A^s_{\mu_4 \mu_3 \mu_2 \mu_1}(s,t),
\end{align}
where the $t \to s$-channel rotation angles are given by
\begin{subequations}
\begin{align}
\cos \chi^{t \to s}_i &=  \frac{(-1)^{c_i+1}(s+m_i^2-m_j^2)(t+m_i^2 - m_k^2) - 2 m_i^2 \Delta_m}{S_{ij}(s) T_{ik}(t)}, \\
\sin \chi^{t \to s}_i &= \frac{2 m_i \phi^{1/2}}{S_{ij}(s) T_{ik}(t)},  \\
\Delta_m &= m_2^2 - m_4^2 - m_1^2 + m_3^2, \\
T^2_{ik}(t) &= \left[t - (m_i + m_k)^2\right]\left[t - (m_i - m_k)^2\right].
\end{align}
\end{subequations}
\end{widetext}
Here, $\phi$ is the Kibble function, $c_i = 1$ ($=0$) if particle $i$ is (not) crossed, $j$ is the $s$-channel and $k$ the $t$-channel pair particle of $i$. Note that, following the path of Trueman and Wick~\cite{Trueman:1964zzb}, the square roots in the $s$-channel residues must be evaluated at $t = t-i\epsilon$, since we cross the real $s,t$ plane at negative $t$. This means that $\sqrt{-t} = i \sqrt{t}$.
Note that we do not include a helicity dependent phase in Eq.~\eqref{eq:crossing_matrix}. This is in agreement with Trueman and Wick~\cite{Trueman:1964zzb}. At leading $s$, we find that
\begin{align}\label{eq:crossing_angles_leading_s}
S_{ij}(s) &= s \\
\sqrt{\phi} &= s\, \sqrt{-t} \\
\sin \chi_i^{t\to s} &= \frac{2 m_i \sqrt{-t}}{T_{ik}(t)} \\
\cos \chi_i^{t\to s} &= \frac{(-1)^{c_i+1} (t+m_i^2-m_k^2)}{T_{ik}(t)}.
\end{align}
The rotation matrix has the property of being factorizable in a top-vertex and bottom-vertex rotation. If we assume for the moment that the left-hand side of Eq.~\eqref{eq:crossing_matrix} can also be factorized
\begin{align}
A^t_{\lambda_4 \lambda_2 \lambda_3 \lambda_1}(s,t) = A^t_{\lambda_3 \lambda_1}(s,t) A^t_{\lambda_4 \lambda_2}(s,t),
\end{align}
then all of the above can be written in a factorized form
\begin{widetext}
\begin{align}
A^t_{\lambda_k \lambda_i}(s,t) &=\sum_{\mu_i, \mu_k} \sqrt{\mathcal{F}(s,t)} \beta^{eik}_{\mu_i \mu_k}(t)  d^{s_i}_{\mu_i \lambda_i}(-\chi^{t \to s}_i) d^{s_k}_{\mu_k \lambda_k}(-\chi^{t \to s}_k) ,
\end{align}
where $(i,k) = (1,3)$ or $(2,4)$. An additional factor of $i$ must be included for the fermion vertex.

In order to write Eq.~\eqref{eq:pwe_t_ch} in a factorized form, we realize that we are working in the high $s$ limit, where $z_t$ is large. The $d$-functions become factorizable when only their leading order in $z_t$ is considered
\begin{align}
A^t_{\lambda_4 \lambda_2 \lambda_3 \lambda_1}(s,t) 
&= 16 \pi \sum_{J = M_t} (2 J + 1) A^t_{\lambda_4 \lambda_2 \lambda_3 \lambda_1, J}(t) e^{-i\pi\lambda'} d^J_{\lambda}(z_t) d^J_{\lambda'}(z_t),
\end{align}
\end{widetext}
where
\begin{align}\label{eq:one_factor_t_ch_d_func}
d^J_{\lambda}(z) \equiv e^{+i\pi\lambda/2}\left[ \left(\frac{z}{2}\right)^{J} \frac{\Gamma(2 J + 1)}{\Gamma(J+\abs{\lambda}+1)\Gamma(J-\abs{\lambda}+1)} \right]^{1/2}.
\end{align}
Note that the above form is not fully factorized in the sense that the $z_t$ depends on both top and bottom particle masses. We therefore consider the leading $s$ form of $z_t$ at constant $t$
\begin{align}
z_t &= \frac{2 t s}{T_{13}(t)T_{24}(t)},
\end{align}
and introduce the functions
\begin{widetext}
\begin{align}\label{eq:one_factor_t_ch_d_func_new}
h^J_{\lambda_{ik}}(z_t) \equiv e^{+i\pi\lambda_{ik}/2}\left[  \frac{\Gamma(2 J + 1)}{\Gamma(J+\abs{\lambda_{ik}}+1)\Gamma(J-\abs{\lambda_{ik}}+1)} \right]^{1/2} \left(\frac{t s}{T_{ik}^2(t)}\right)^{J/2}.
\end{align}
\end{widetext}

Close to the $m_e$ pole in $t$, the partial-wave expansion is dominated by the $J_e$ partial wave. In Eq.~\eqref{eq:pw_pole_tch} we have assumed a factorizable form for the $t$-channel partial-wave amplitude. Therefore, we can rewrite the partial-wave amplitude as 
\begin{align}
A^t_{\lambda_4 \lambda_2 \lambda_3 \lambda_1, J_e}(t) &= A^t_{\lambda_4 \lambda_2, J_e}(t) A^t_{\lambda_3 \lambda_1, J_e}(t),
\end{align}
where 
\begin{align}\label{eq:one_factor_t_ch_pw_amp}
A^t_{\lambda_k \lambda_i, J_e}(t) &= \gamma_{\lambda_i \lambda_k}(t) \sqrt{\frac{1}{t-m_e^2}}.
\end{align}
The above can then be written in the factorized form
\begin{align}\label{eq:one_factor_t_ch_pw_amp_new}
A^t_{\lambda_k \lambda_i}(s,t) &= \sqrt{16 \pi (2 J_e + 1) } A^t_{\lambda_k \lambda_i, J_e}(t) h^{J_e}_{\lambda_{ik}}(z_t) \xi(\lambda_{ik}),
\end{align}
where
\begin{align}
\xi(\lambda_{ik}) = e^{-i\pi \lambda_{ik}} \text{ for a bottom vertex, else } \xi(\lambda_{ik}) =1.
\end{align}
Putting everything together, we obtain the explicit form of the $t$-channel residue as a function of the $s$-channel residue
\begin{widetext}
\begin{align}
 \gamma_{ \lambda_i \lambda_k}(t) &= \frac{1}{\sqrt{16 \pi (2 J_e + 1)}} \frac{s^{J_e/2}}{\xi(\lambda_{ik}) h^{J_e}_{\lambda_{ik}}(z_t)} \sum_{\mu_i, \mu_k}   \beta^{eik}_{\mu_i \mu_k}(t)  d^{s_i}_{\mu_i \lambda_i}(-\chi^{t \to s}_i) d^{s_k}_{\mu_k \lambda_k}(-\chi^{t \to s}_k).
\end{align}
Note that the above is only valid for $s \to \infty$ and $t \to m_e^2$.
In order to explicitly illustrate the cancellation of the $s$-dependence, we focus on the $h^{J_e}_{\lambda_{ik}}(z_t)$ function.
Using Eq.~\eqref{eq:one_factor_t_ch_d_func_new}, one obtains
\begin{align}
\frac{h^{J_e}_{\lambda_{ik}}(z_t)}{s^{J_e/2}} &= e^{+i\lambda_{ik}/2} \left(\frac{t}{T_{ik}^2(t)}\right)^{J_e/2} \left[ \frac{\Gamma(2 J_e + 1)}{\Gamma(J_e+\abs{\lambda_{ik}}+1)\Gamma(J_e-\abs{\lambda_{ik}}+1)} \right]^{1/2}.
\end{align}
Hence (and being more precise in the notation)
\begin{align}\label{eq:relation_s_t_residues_top}
 \gamma^e_{\lambda_1 \lambda_3}(t = m_e^2) =& \frac{e^{-i\pi \lambda_{13}/2} }{\sqrt{16 \pi (2 J_e + 1)}} 
 \left(\frac{T_{13}(m_e^2)}{m_e}\right)^{J_e} \left[ \frac{\Gamma(J_e+\abs{\lambda_{13}}+1)\Gamma(J_e-\abs{\lambda_{13}}+1)}{\Gamma(2 J_e + 1)} \right]^{1/2} \nonumber \\
&\times \sum_{\mu_1, \mu_3}   \beta_{\mu_1 \mu_3}(t = m_e^2)  d^{s_1}_{\mu_1 \lambda_1}(-\chi^{t \to s}_1\vert_{t=m_e^2}) d^{s_3}_{\mu_3 \lambda_3}(-\chi^{t \to s}_3\vert_{t=m_e^2}) ,
\end{align}
for the top vertex and 
\begin{align}\label{eq:relation_s_t_residues_bottom}
 \gamma^e_{\lambda_2 \lambda_4}(t = m_e^2) =& \frac{(-i) e^{+i\pi \lambda_{24}/2} }{\sqrt{16 \pi (2 J_e + 1)}} 
 \left(\frac{T_{24}(m_e^2)}{m_e}\right)^{J_e} \left[ \frac{\Gamma(J_e+\abs{\lambda_{24}}+1)\Gamma(J_e-\abs{\lambda_{24}}+1)}{\Gamma(2 J_e + 1)} \right]^{1/2} \nonumber \\
&\times \sum_{\mu_2, \mu_4}   \beta_{\mu_2 \mu_4}(t = m_e^2)  d^{s_2}_{\mu_2 \lambda_2}(-\chi^{t \to s}_2\vert_{t=m_e^2}) d^{s_4}_{\mu_4 \lambda_4}(-\chi^{t \to s}_4\vert_{t=m_e^2}) ,
\end{align}
for the bottom vertex.
\end{widetext}
The crossing angles must be evaluated at the pole
\begin{align}
\sin \chi_i\vert_{t=m_e^2} &= i\frac{2 m_i  m_e}{\lambda^{1/2}(m_e^2, m_i^2, m_k^2)} , \\
\cos \chi_i\vert_{t=m_e^2} &= \frac{(-1)^{c_i+1} (m_e^2+m_i^2-m_k^2)}{\lambda^{1/2}(m_e^2, m_i^2, m_k^2)}.
\end{align}
In summary, Eqs.~\eqref{eq:relation_s_t_residues_top} and \eqref{eq:relation_s_t_residues_bottom} relate the $s$-channel residues at the pole directly to the $t$-channel residues. With these expressions, one can compare the results obtained in this work directly to the decay couplings of various processes.

\newpage
\bibliographystyle{apsrev4-1-jpac}
\bibliography{quattro.bib}

\end{document}

%% file: jpac.tex
\usepackage{graphicx}
\usepackage{dcolumn}
\usepackage{bm}
\usepackage{hyperref}
\usepackage{cleveref}
\usepackage{dsfont}
\usepackage{placeins}
\usepackage{todonotes}
\usepackage[utf8]{inputenc}
\usepackage{array}   
\newcolumntype{L}{>{$}l<{$}} 
\newcolumntype{R}{>{$}r<{$}}
\newcolumntype{C}{>{$}c<{$}}
\usepackage{color}
\usepackage{xcolor}
\usepackage{xspace}
\usepackage{braket}
\usepackage{units}
\usepackage{slashed}
\usepackage[normalem]{ulem}
\usepackage[format=plain,justification=RaggedRight,singlelinecheck=false]{caption}
\usepackage[format=plain,justification=centering,singlelinecheck=false]{subcaption}

\newcommand{\abs}[1]{\ensuremath{|#1|}}

\newcommand{\propagator}[1]{\mathcal{P}_{#1}}
\newcommand{\reggepropagator}[1]{\mathcal{F}_{#1}}
\newcommand{\pomeron}{\mathbb{P}}
\newcommand{\nn}{\nonumber}

\newcommand{\plab}{{\ensuremath {p_\text{lab}}}}
\newcommand{\kbarz}{\ensuremath{\overline{K^0}}}
\newcommand{\betahat}{\ensuremath{\hat\beta}}

\renewcommand{\Im}{\ensuremath{\textrm{Im }}}

\newcommand{\eg}{{\it e.g.}\xspace}
\newcommand{\etal}{{\it et al.}\xspace}
\newcommand{\ie}{{\it i.e.}\xspace}
\newcommand{\unitm}{\ensuremath{\mathds{1}}\xspace}

\newcommand{\mev}{\ensuremath{{\mathrm{\,Me\kern -0.1em V}}}\xspace}
\newcommand{\gev}{\ensuremath{{\mathrm{\,Ge\kern -0.1em V}}}\xspace}
\newcommand{\tev}{\ensuremath{{\mathrm{\,Te\kern -0.1em V}}}\xspace}
\newcommand{\diffd}{\ensuremath{{\mathrm{d}}}}

\usepackage{mfirstuc} 
\newcommand{\addReviewer}[2]{
  \expandafter\newcommand\csname #1\endcsname[1]{{\bf \color{#2} \capitalisewords{#1}:\,##1}}
  \expandafter\newcommand\csname #1cor\endcsname[2]{{\color{#2} \capitalisewords{#1}:\,\st{##1}{\bf ##2}}}
  \expandafter\newcommand\csname #1color\endcsname{#2}
}

\usepackage{tikz, marginnote} 
\newcommand{\checkedby}[1]{
\ifdefined\CROSSCHECKS
  \marginnote{
    \begin{tikzpicture}
      \foreach \x [count=\xi] in {#1} {
         \node[shape=circle,inner sep=0mm,
         minimum size=2mm,
         fill=\csname \x color\endcsname] at (\xi*3mm,0) {};
       }
    \end{tikzpicture}
  }
\else
\fi
}

\usepackage{soul,color}
\definecolor{chromeyellow}{rgb}{1.0, 0.65, 0.0}
\definecolor{DodgeBlue}{rgb}{0.118, 0.565,1.000}
\definecolor{asparagus}{rgb}{0.53, 0.66, 0.42}
\definecolor{cadmiumgreen}{rgb}{0.0, 0.42, 0.24}

\addReviewer{adam}{red}
\addReviewer{ale}{chromeyellow}
\addReviewer{vincent}{brown}
\addReviewer{misha}{cadmiumgreen}
\addReviewer{jannes}{blue}
\addReviewer{cesar}{magenta}
\addReviewer{astrid}{cyan}
\addReviewer{miguel}{DodgeBlue}
\addReviewer{andrew}{purple}
\addReviewer{nathan}{orange}
\addReviewer{arkaitz}{asparagus}